\documentclass[fleqn,usenatbib]{mnras}

\usepackage{newtxtext,newtxmath}

\usepackage[T1]{fontenc}
\usepackage[utf8]{inputenc}

\DeclareRobustCommand{\VAN}[3]{#2}
\let\VANthebibliography\thebibliography
\def\thebibliography{\DeclareRobustCommand{\VAN}[3]{##3}\VANthebibliography}

\usepackage{graphicx}	
\usepackage{amsmath}	
\usepackage{multirow, blindtext}
\usepackage{soul}
\usepackage{subfig,float}

\defcitealias{Fontanot+20}{F20}
\defcitealias{Ueda+14}{U14}
\defcitealias{Buchner+15B}{B15}
\defcitealias{Ananna+19}{A19}
\defcitealias{Alonso-Tetilla}{Paper I}
\defcitealias{Wada15}{Wada}
\defcitealias{Ramos-Almeida+17}{RA\&R}
\DeclareUnicodeCharacter{2212}{\textendash}
\usepackage{orcidlink}
\DeclareUnicodeCharacter{02BC}{'}

\title[AGN light curves and obscuration]{{Characterizing the roles of transitory obscured phases and inner torus in shaping the fractions of obscured AGN at cosmic noon}
}

\author[A. V. Alonso-Tetilla et al.,]{Alba V. Alonso-Tetilla$^{\orcidlink{0000-0002-6916-9133},1}$\thanks{E-mail: A.V.Alonso-Tetilla@soton.ac.uk},
Francesco Shankar$^{\orcidlink{0000-0001-8973-5051},1}$\thanks{E-mail: F.Shankar@soton.ac.uk},
Fabio Fontanot$^{2,}$$^{3}$,
Andrea Lapi$^{4}$, \newauthor
Milena Valentini$^{5,}$$^{2,}$$^{6,}$$^{3,}$$^{7}$,
Annagrazia Puglisi$^{1,}$\thanks{Anniversary Fellow},
Nicola Menci$^{8}$,
Hao Fu$^{9,1}$, 
Lumen Boco$^{4}$, \newauthor
Johannes Buchner$^{10}$, 
Michaela Hirschmann$^{11,}$$^{2}$, 
Cristina Ramos Almeida$^{12,}$$^{13}$, 
Carolin Villforth$^{14}$, 
Lizhi Xie$^{15,}$$^{2}$ 
\\
$^{1}$ School of Physics and Astronomy, University of Southampton, Highfield, SO17 1BJ, Southampton, UK\\
$^{2}$ INAF - Astronomical Observatory of Trieste, via G.B. Tiepolo 11, I-34143 Trieste, Italy\\
$^{3}$ IFPU – Institute for Fundamental Physics of the Universe, via Beirut 2, I-34151 Trieste, Italy \\
$^{4}$ SISSA, Via Bonomea 265, 34136 Trieste, Italy \\
$^{5}$ Department of Physics of the University of Trieste, Astronomy Section, via Tiepolo 11, I-34131 Trieste, Italy  \\
$^{6}$ ICSC - Italian Research Center on High Performance Computing, Big Data and Quantum Computing, via Magnanelli 2, 40033, Casalecchio di Reno, Italy \\
$^{7}$ INFN, Instituto Nazionale di Fisica Nucleare, Via Valerio 2, I-34127, Trieste, Italy \\
$^{8}$ INAF - Osservatorio Astronomico di Roma, via Frascati 33, I-00078 Monteporzio, Italy \\
$^{9}$ Center for Astronomy and Astrophysics and Department of Physics, Fudan University, Shanghai 200438, People’s Republic of China \\
$^{10}$ Max Planck Institute for Extraterrestrial Physics, Giessenbachstrasse, 85741 Garching, Germany \\
$^{11}$ Institute for Physics, Laboratory for galaxy Evolution and Spectral Modelling, EPFL, Observatoire de Sauverny, Chemin Pegasi 51, 1290 Versoix, Switzerland \\
$^{12}$ Instituto de Astrofísica de Canarias, Calle Vía Láctea, s/n, E-38205, La Laguna, Tenerife, Spain \\
$^{13}$ Departamento de Astrofísica, Universidad de La Laguna, E-38206, La Laguna, Tenerife, Spain \\
$^{14}$ Department of Physics, University of Bath, Claverton Down, Bath BA2 7AY, UK \\
$^{15}$ Tianjin Normal University, Binshuixidao 393, Xiqing, 300387, Tianjin, Peopleʼs Republic of China 
}

\date{Accepted 2025 August 15. Received 2025 August 14; in original form 2024 September 06 }

\pubyear{\today}

\begin{document}
\maketitle

\begin{abstract}
The origin of obscuration in Active Galactic Nuclei (AGN) is still a matter of contention. It is unclear whether obscured AGN are primarily due to line-of-sight effects (Orientation model), a transitory, dust-enshrouded phase in galaxy evolution (Evolution models), or a combination of both. The role of an inner torus around the central supermassive black hole also remains unclear in pure Evolution models. We use cosmological semi-analytic models and semi-empirical prescriptions to explore obscuration effects in AGN {at cosmic noon, in the range $1<z<3$}. We consider a realistic object-by-object modelling of AGN evolution including different AGN light curves {(LCs)} composed of phases of varying levels of obscuration, {usually (but not uniquely) with a larger degree of obscuration before the peak of AGN activity, mimicking the possible clearing effects of strong AGN feedback.} Evolution models {characterized by AGN LCs with relatively short pre-peak obscured phases followed by more extended optical/UV visible post-peak phases,} struggle to reproduce {the high fraction of obscured AGN} at $z \sim 2-3$ inferred from X-ray surveys. Evolution models {characterised by AGN LCs with} sharp post-peak declines or persistent or multiple obscuration phases are more successful, {although they still face challenges in reproducing the steady drop in the fractions of obscured AGN with increasing luminosity measured by some groups. Invoking a fine-tuning in the input LCs, with more luminous AGN defined by longer optical/UV visible windows, can improve the match to the decreasing fractions of obscured AGN with luminosity. Alternatively, a long-lived central torus-like component, with thickness decreasing with increasing AGN power, naturally boosts the luminosity-dependent fractions of obscured AGN, suggesting that small-scale orientation effects may still represent a key component even in Evolution models. We also find that in our models major mergers and starbursts, when considered in isolation, fall short in accounting for the large fractions of highly obscured faint AGN detected at cosmic noon.} 
\end{abstract}




\section{Introduction}

Actively accreting Supermassive Black Holes (SMBHs) at the centre of massive galaxies, i.e., Active Galactic Nuclei (AGN), are considered obscured when the emission from the accretion disc at X-ray, ultraviolet, and optical wavelengths is absorbed by material along the line-of-sight \citep[e.g.,][]{Seyfert43, Antonucci93, Urry_Padovani95, Netzer15, Ramos-Almeida+17}. 

Dissecting the origin of obscuration is pivotal in determining a complete census of AGN and, in turn, in providing a more complete description of the evolution of SMBHs in a cosmological context \citep[e.g.,][]{Shankar+13, Hickox+18, Yutani+22, Ricci+22, Petter+23}. Models and observations propose that both dust and gas in the interstellar medium of the host galaxy \citep[][and references therein]{Lapi+05, Buchner+17, Gilli+22, Andonie+23} along with a dusty torus located at a few parsecs from the central SMBH \citep[e.g.,][]{Packham+05, Radomski+08, Burtscher+13, Gallimore+16, Garcia-Burillo+16, Imanishi+16, Garcia-Burillo+21, Natalie25}, are viable sources of AGN obscuration. The obscuration from the host galaxy is expected to vary according to its gas and dust content (and distribution), and is expected to be more prominent during the early phases of host galaxy formation \citep[$z\sim1.5$, e.g.,][]{Granato+04, Santini+14, Lapi+06, Lapi+14}. 

The torus was originally thought to be a long-lived structure \citep[][]{Urry_Padovani95}, but more recent studies suggest that this is a dynamical, clumpy structure created from the accretion disc and part of the dusty wind \citep[e.g.,][]{Ramos-Almeida+09, RamosAlmeida+11, RamosAlmeida16, Wada12, Markowitz+14, Lopez-Gonzaga+16, Ramos-Almeida+17, Honig+17, Hoenig19}. Given the complex interplay between the host galaxy interstellar medium evolution and the physics and evolution of gas accreting onto the central black hole, studying the AGN obscuration can provide a more complete understanding of SMBH physics and demographics, and set important constraints on how they (co-)evolve with their host galaxies across cosmic time \citep[e.g.,][]{Shankar+13,Ricci+22, Yutani+22, Petter+23}.

Currently, there are two leading theories that can explain the obscuration in AGN, and these are not mutually exclusive. {\textit{Orientation} (also called \textit{Unification}) models assume that the gas and dust distributions within the host galaxy and a central dusty torus are the primary sources of AGN obscuration \citep[e.g, ][]{Antonucci93, Urry_Padovani95, Polletta+99, Netzer15, Alonso-Tetilla}, with the level of obscuration depending on the inclination of the system relative to the observer's line of sight and the torus covering factor \citep[e.g.,][]{RamosAlmeida+11}}. \textit{Evolutionary} models propose {instead} that the level of obscuration, as well as the AGN luminosity and {SMBH} mass, {depend on the specific time when the AGN is observed within the light curve (LC),} regardless of {any line} of sight \citep[e.g, ][]{Sanders+89, Granato+04, Lapi+06, Hopkins+07, Shankar08BAL}. {In this second class of models, obscuration is mostly contributed by \textit{transitory, short-term,} and early dust-enshrouded phases of the host galaxy, though a contribution from an inner torus could still be present, as also further explored in this work.} 
This \textit{pure} Evolution model proposes a \textit{two-mode} growth for the central SMBH, with a first obscured pre-peak (super-)Eddington phase, {analytically described by an exponential increase in the SMBH accretion rate, usually occurring during a gas-rich star-forming phase in the host galaxy,} followed by a non-obscured phase, {described by a power-law decline \citep[e.g.,][]{Shen09,Paolillo+22}}. {The} post-peak phase is expected to be more extended, corresponding to sub-Eddington accretion levels onto the central SMBH, which is massive enough to be able to clear out and/or ionize the host galaxy interstellar medium with winds and/or jets \citep[e.g., reviews by][]{ Brandt_Hasinger05, AlexanderHickox12, SomervilleDave}, and reveal the AGN at X-ray, UV, and optical wavelengths. {Some theoretical models suggest that pure Evolution models, with the pre-peak, exponential phase of the LC heavily obscured, may be able to reproduce the luminosity function (LF) of obscured vs unobscured AGN without the need for an inner torus,} although a clear consensus on this key aspect has yet to be reached \citep[e.g.,][]{Granato+06, Lapi+06, Hopkins+08, Lapi+14, Hickox+20, Viitanen+23, Georgantopoulos+23}.

Observations indicate that both orientation and evolutionary effects can play a role in obscuring the AGN \citep[e.g.,][]{Polletta+08, Polletta+11}. For example, {in the original definition of} \textit{Type 2} AGN \citep[for definitions and \textit{Type} differences see][]{Antonucci93}{, those galaxies} lack broad emission lines. This characteristic is usually interpreted as an orientation effect since broad emission lines are believed to originate in the inner clouds orbiting the central black hole, which, assuming typical covering factors of $\sim 20-40\%$ \citep[][]{Khachikian+74, Peterson+1997, Kuraszkiewicz+21, Kuhn+24}, should remain invisible to edge-on lines of sight intersecting the torus aperture. {Indeed, in support of the Orientation model, detections of broad lines in polarised light have been found, indicating the presence of a hidden broad line region in Type 2 AGN \citep[][]{Barth+99, Nagao+04, Yu+05, RamosAlmeida+08}.} On the other hand, galaxies hosting \textit{Type 1} AGN appear in several instances more massive than \textit{Type 2} AGN, at odds with expectations from basic Unification models \citep{RicciF+22}. In a similar vein, some studies suggest that AGN are ubiquitous in starburst and actively star-forming galaxies at different redshifts \citep[e.g.,][and references therein]{Alexander+05, Alexander+16, Rodighiero+15, Mullaney+15, Mountrichas+23, Mountrichas+24}, in line with predictions from Evolution models.

An evolutionary connection between star formation, black hole accretion and fuelling \citep[\textbf{e.g.,}][]{Granato+06, Shankar+04, Shankar+12, Shankar+13}, and AGN obscuration is expected given their common dependence on the cold gas content of the host galaxy itself \citep[e.g.,][]{Harrison17}. In the context of Evolution models, in fact, many works have identified the presence of galaxies with high cold gas fractions and star formation rates (SFRs) in the obscured AGN population because of, e.g., high levels of dust \citep[e.g.,][]{Afonso+03, Wijesinghe+11, Chen+15} as well as carrying the signatures of relatively recent mergers \citep[e.g.,][]{Darg+10, RicciC+17b, Yutani+22, Pierce+23}. Starburst and enhanced SFRs are often the result of mergers, disc instabilities, and interactions \citep[e.g., from observations, \citealt{Armus+87, Kennicutt+87, Xu+91, Sanders+96, Ellison+13, Knapen+15, Silva+21, Cezar+24}, and e.g., from simulations][, and references therein]{Hernquist+89, Springel+05, DiMatteo+08, Teyssier+10, Moreno+15, Renaud+22, Linden+22}, further supporting the presence of a strong link between mergers and evolutionary effect in obscured AGN \citep{Andonie+23}.

Despite the variety of methods to characterize obscured AGN \citep[for an extensive overview see][]{Hickox+18}, in this paper we focus on X-ray observations, since they are directly associated with the accretion discs and its hot corona \citep[e.g.,][]{Giacconi09}, and have more penetrating power through thick column densities. Several studies have tried to quantify the level of obscuration in AGN through the use of HI-equivalent $N_{\rm H}$ column densities \citep[e.g.,][]{Ueda+14, Aird+15, Buchner+15B, Ananna+19, Brivael22}, and we will follow the same methodology in the present work when comparing our models to observations. This paper is the second of a series that explores AGN obscuration in the context of the semi-analytic model (SAM) for GAlaxy Evolution and Assembly \citep[GAEA,][\citetalias{Fontanot+20} hereafter]{Fontanot+20}. This theoretical model provides realistic properties of galaxies and their central black holes, whilst providing detailed predictions on many key properties such as the SMBH accretion rate and SFR distributions, and AGN bolometric Light Curves (LCs). The flexibility of GAEA allows us to explore the impact on our predictions of varying the input AGN LC, gas fractions, galaxy structure, as well as exploring the connection between AGN obscuration, starburst and major mergers. In the first paper of this series \citep[][\citetalias{Alonso-Tetilla} hereafter]{Alonso-Tetilla}, we explored the impact of orientation only and found that {galaxy-scale obscuration, at least when described via an exponential gas disc, can only account for part of the Compton-thin obscuration, while adding a central torus component can boost the obscured fractions of both Compton-thin and Compton-thick AGN.} 

{In this paper we go a step further and explore model with Evolutionary and Orientation features.} {In pure Evolutionary models}, we bypass the details of the geometry and/or the amount of the cold gas entirely, {and assume} that the level of $N_{\rm H}$ reached by a single galaxy is only dictated by the position of the AGN within the LC, being, for example, Compton-thick only for a fraction of the time during the pre-peak phase. {We will complement our study with a composite class of Evolution+Orientation models in which we still retain the idea that the AGN is obscured or visible during specific, pre-defined phases within the LC, but we also include elements proper of an Orientation model. More specifically, 1) we first include an inner torus and investigate the conditions under which a long-lived, small-scale obcuring structure can contribute to the fraction of deeply buried AGN at cosmic noon; 2) we then also explore the impact of assigning an $N_{\rm H}$ column density during the obscured phases within the LC not at random, but based on the actual gas mass in the host galaxy as carried out in \citetalias{Alonso-Tetilla}.}

The structure of this paper is as follows. In Section \ref{sec:data} we review the theoretical models and introduce the observational data used throughout the paper. In Section \ref{sec:methodology} we describe our methodology in the framework of the GAEA SAM, and we provide the details of our AGN and host galaxy {obscuration} modelling. In Section \ref{sec:results} we present our predicted fractions of obscured AGN in the context of our Evolution and Orientation models. {In Section \ref{sec:discussion} we discuss our main results}. Finally, in Section \ref{sec:conclusions} we summarize our conclusions.

\section{Theoretical models, observational data, and comparison strategy}
\label{sec:data}

In this Section we describe {the semi-analytic models and coupled semi-empirical prescriptions we adopt} to predict the distribution of obscured Compton-thin (CTN) and Compton-thick (CTK) AGN under the Evolution {and Orientation frameworks}. {Our reference background galaxy model is the semi-analytic model GAEA (Section~\ref{sec:GAEA}), which provides a realistic representation of galaxies in the mass and redshift range of interest here. However, we stress that our analysis is independent of the specific underlying reference galaxy evolution model, and indeed we will consider the impact of varying input LCs, gas fractions, or other relevant galactic properties.} 

\subsection{The GAEA semi-analytic model}  \label{sec:GAEA}

{The state-of-the-art semi-analytic model GAEA follows the evolution of galaxies and their central SMBHs in a cosmological volume from early times to the present epoch}. GAEA includes prescriptions {for} the evolution of the baryonic component in galaxies, as well as providing a detailed modelling of the growth of the central SMBHs. In the following, we summarize the key ingredients characterizing the modelling of the cold gas accretion onto SMBHs in GAEA, and refer the reader to \citetalias{Fontanot+20} for a more detailed description. In particular, in this paper we focus on the so-called HQ11-GAEA model, which includes the \citet{Hopkins&Quataert2011} and \citet{Hopkins+06} prescriptions to estimate:
\begin{enumerate} 
    \item The fraction of cold gas from the host galaxy that, {after a merger or disc instability, is able} to lose {sufficient} angular momentum to reach the central regions and gather onto a low angular momentum gas reservoir. 
    \item {The mass accretion rate onto the SMBH from the low angular momentum reservoir, which follows the shape of a LC inspired by the results of hydrodynamic simulations (see Section~\ref{subsec:LCs}).} 
\end{enumerate}

SMBH seeding in GAEA is performed following \citet{Volonteri+11} which implies seed masses of $\sim 10^4$ M$_\odot$. {The seeds grow from cold gas accretion, triggered by galaxy mergers or disc instabilities, and mergers with other SMBHs. The energy arising from the accretion is redistributed in time following an AGN LC, composed of an initial super-Eddington accretion phase, which lasts until the SMBH reaches the self-regulation limit \citep{Hopkins+06}, followed by a power-law decline.}
{Similar behaviours for the AGN LC} have been suggested by theoretical arguments, hydrodynamic simulations \citep[e.g.,][]{Granato+04, Lapi+06, Hopkins+07, Shen09} and phenomenological models \citep[e.g.,][]{Lapi+06, Lapi+14, Aversa+15}. GAEA also includes QSO-mode feedback in the form of AGN driven outflows, {AGN winds capable of heating the cold gas and eventually returning it into the hot phase}. Specifically, the model realization considered in this work, HQ11-GAEA, uses the outflow rate predictions as a function of cold gas mass, bolometric luminosity and black hole mass from \citet{Menci+2019}.

The original GAEA model has been calibrated on Dark Matter Merger trees drawn from the Millennium Simulation \citep[][WMAP1 lambda cold dark matter concordance cosmology, i.e., $\Omega_{\Lambda}=0.75$, $\Omega_{\rm m}=0.25$, $\Omega_{b} = 0.045$, $n=1$, $\sigma_8 = 0.9$, and $H_0=73$ km s$^{-1}$ Mpc$^{-1}$]{Springel+05}, {and} yields a good description of galaxy properties down to a stellar mass scale of $\sim 10^9$ M$_{\odot}$. The HQ11-GAEA model is calibrated to reproduce the evolution of the AGN\footnote{We define as AGN those galaxies shining above $L_{\rm bol} \sim 10^{42}$ erg/s at some point during their active phase.} bolometric LF up to redshift $z\sim4$, i.e., without applying any obscuration correction to model predictions, and it also reproduces all galaxy properties discussed in previous papers \citep[e.g.,][]{Hirschmann+16}, like mass-metallicity relations, quenched fractions and cold gas fractions. The HQ11-GAEA model also reproduces the observed Eddington ratio distribution function at various redshifts \citepalias{Fontanot+20}. A deeper analysis on the chemical enrichment can be found in \citet{DeLucia+14}.

\begin{figure*}
    \centering
    \includegraphics[width=0.9\textwidth]{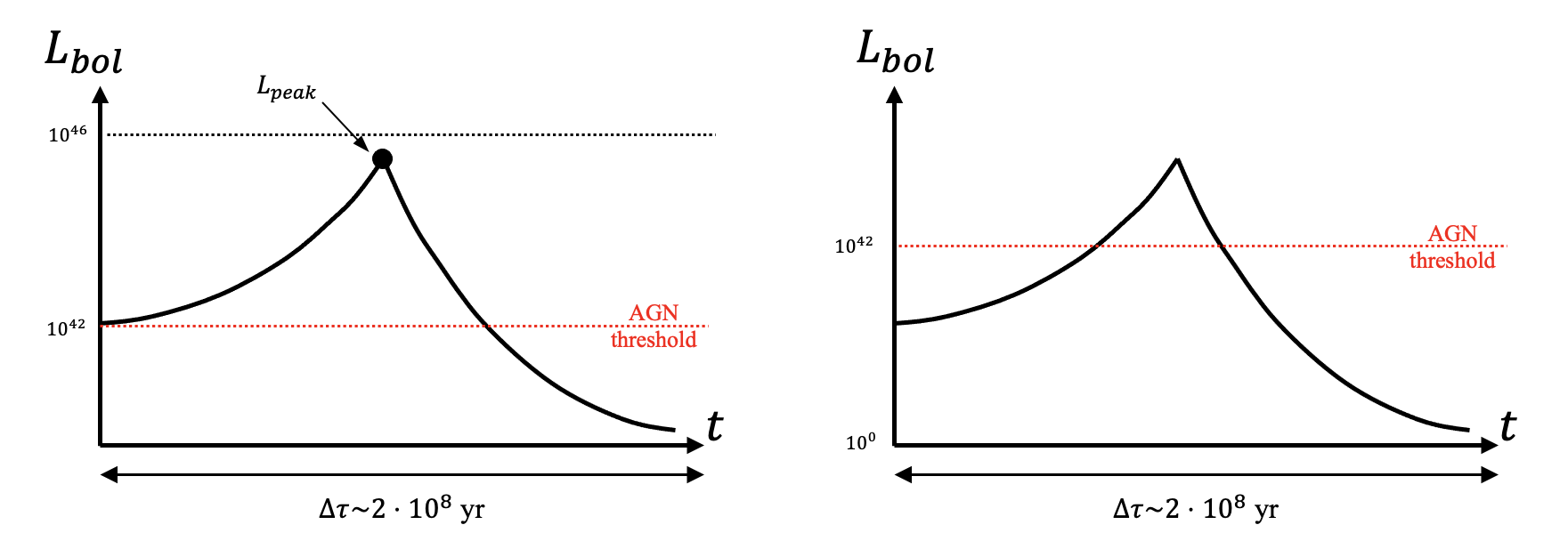}
    \caption{
    {General sketch of the Light Curves (LCs) used as a baseline in this work, comprising of an exponential pre-peak growth phase followed by a more extended sub-Eddington phase (which is not included in some model renditions, see Figure \ref{fig:models14}), {and with a maximum temporal extent of $\Delta \tau\sim 2\cdot 10^8$ yr} at $z=2.4$, as described in Section \ref{subsec:LCs}. We show two noteworthy examples of LCs with a peak luminosity well above (left panel) the minimum bolometric luminosity considered in this work (horizontal dotted, red lines), allowing for the whole LC to be well sampled, and one LC with a peak luminosity just slightly above (right panel) the minimum luminosity. Bolometric luminosities are in units of erg/s.}
    }
    \label{fig:SketchBaselineLC}
\end{figure*}

{GAEA also includes prescriptions for the radio-mode accretion and feedback}. The radio mode is treated, by construction, as an (almost) continuous accretion process of hot gas from the halo \citep[which gives rise to tensions with the observed distribution of radio galaxies - see e.g.,][]{Fontanot+11}. This SMBH accretion mode becomes relevant for massive galaxies residing in massive haloes at low redshifts $z<1$, as those are the environments where efficient quenching of the cooling flows and late SFR are required. 
Since in this paper we mainly focus on $z>1$, {around the peak of AGN activity,} we will neglect the contribution of radio-mode accretion to the bolometric luminosity, similarly to what we {assumed} in \citetalias{Alonso-Tetilla}. This choice has the additional advantage of simplifying the treatment of the LC, focusing on the QSO-mode contribution only. {We also note that the GAEA models are calibrated to reproduce the $z<4$ AGN bolometric LFs, but do not include any explicit treatment for obscuration, which is the focus of the current work}. The 2-10 keV intrinsic X-ray luminosities are calculated from bolometric luminosities via the bolometric correction by \citet{Duras+20}, {although similar results} would be {recoved by} adopting, for example, the \citet{Marconi+04} bolometric correction.

\begin{figure*}
    \centering
    \includegraphics[width=0.9\textwidth]{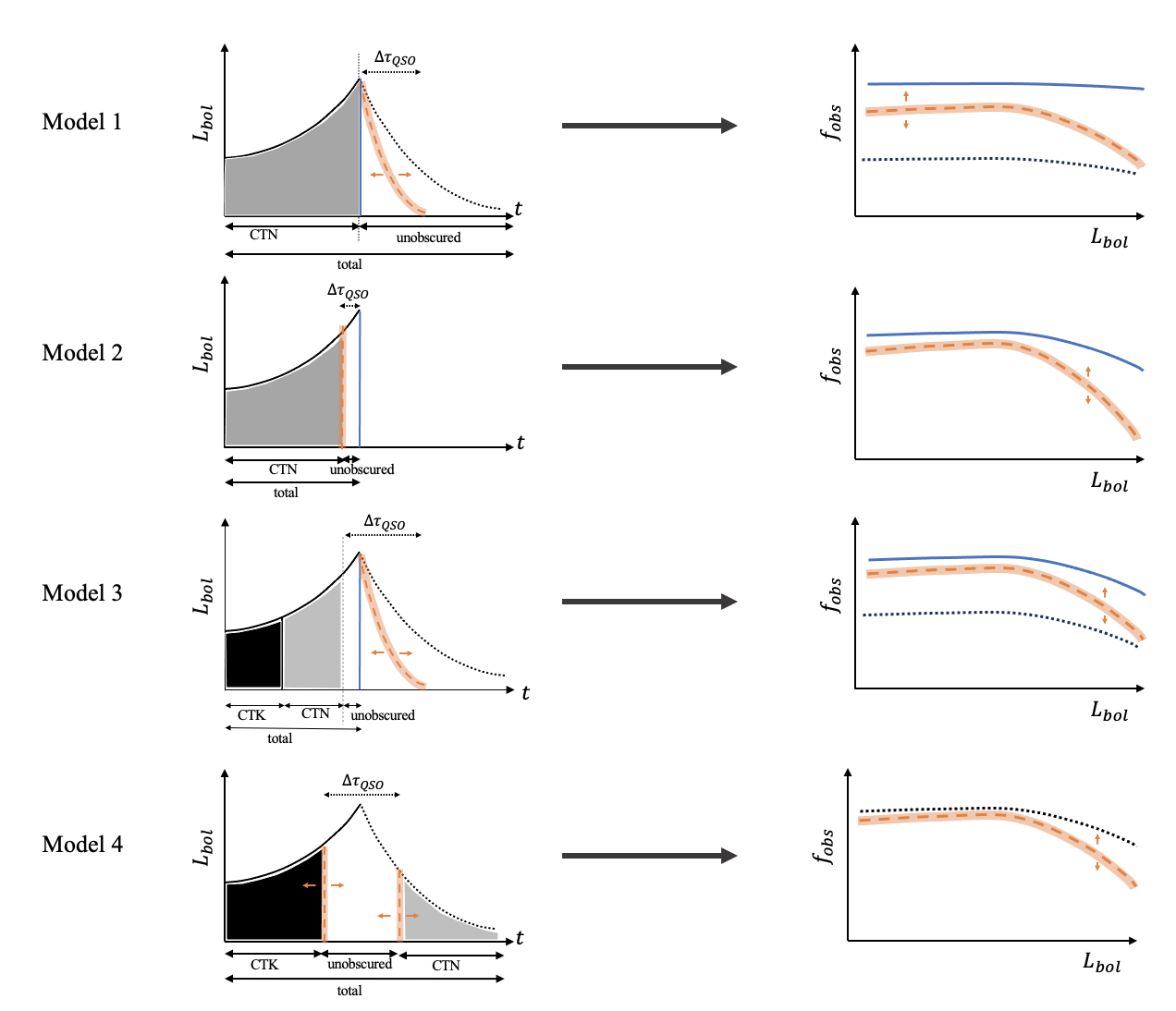}
    \includegraphics[width=0.2\textwidth]{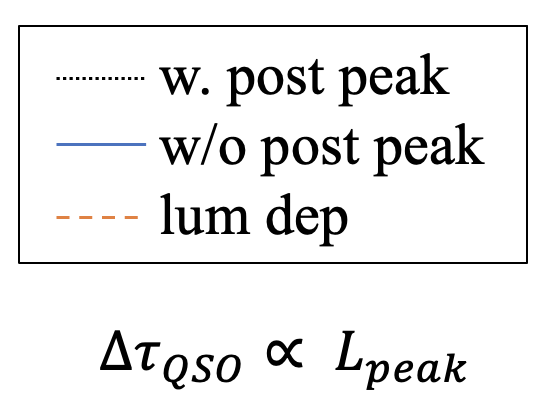}

    \caption{
    {General sketch of the Light Curves (LCs) in our reference Evolutionary Models, including a qualitative definition of the main {obscuration parameters considered} in this study (left column). We also provide the foreseen impact on the predicted fractions of obscured AGN (right column). The grey areas mark the Compton-thin (CTN) phases, while the dark sections mark the Compton-thick (CTK) phases within the light curves. All underlying light curves in each Model are identical in shape, but differ in terms of i) the post-peak behaviour, including a prolonged post-peak phase (black dotted lines) or having a sharp cut-off at the peak (blue solid lines), or ii) for the internal redistribution of the CTN/CTK phases}. {Sharp cut-offs in the light curves after the peak tend to generate larger fractions of obscured AGN (solid blue lines) compared to models with prolonged post-peak phases (dotted lines)}. 
    {We define $\Delta \tau_{\rm QSO}$ as the time of the optically/UV visibility window, {within the maximum temporal extent of the light curve of $\Delta \tau\sim 2\cdot 10^8$ yr} at $z=2.4$ {(Figure \ref{fig:SketchBaselineLC})}, our mean redshift of interest. Models with a luminosity-dependent visibility window $\Delta \tau_{\rm QSO}(L_{\rm peak})$ present a stronger decrease in the obscured fractions of AGN with increasing luminosity (orange dashed lines). Bolometric luminosities are all in units of erg/s.}
    }
    \label{fig:models14}
\end{figure*}%
\begin{figure*}\ContinuedFloat
    \centering
    \includegraphics[width=0.9\textwidth]{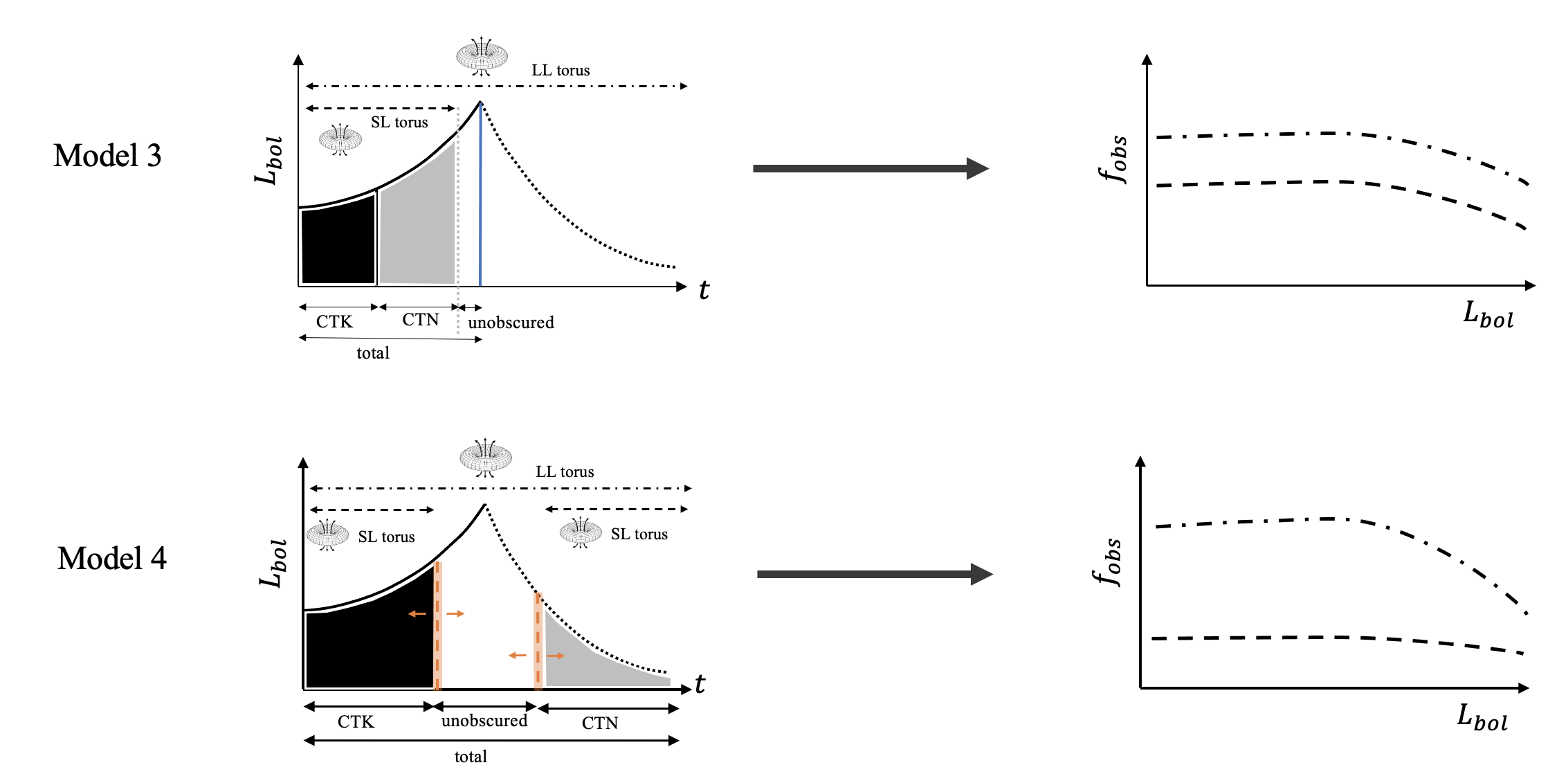}
    \includegraphics[width=0.2\textwidth]{Figures/diagrams/Figure2_new.png}
    \caption[]{ \textbf{(Cont.)}
    {Sketch of the Evolution Models 3 and 4, in the same format as above, highlighting the inclusion of a torus model, assumed to be either short-lived (``SL''; black long-dashed lines) or long-lived (``LL''; black dot-dashed lines), i.e., surviving even during the optical/UV phases within the AGN light curves. A long-lived torus tends to boost the fractions of both CTN/CTK AGN.}
    }
    \label{fig:models14}
\end{figure*}

\subsection{Observational data}\label{sec:obsdata}

{We compare our predicted fractions of obscured AGN as a function of 2–10 keV intrinsic X-ray luminosity with the empirical measurements by \citet[][]{Ueda+14}, \citet[][]{Buchner+15B}, and \citet[][]{Ananna+19} \citepalias[][hereafter]{Ueda+14, Buchner+15B, Ananna+19}}. These observations include data from deep surveys from observatories such as \textit{Swift}/BAT, ASCA, \textit{XMM-Newton}, \textit{Chandra}, or \textit{ROSAT}, being among the most complete compilations in terms of AGN luminosity and redshift coverage. We specifically use the two forms {of AGN obscured fractions} from \citetalias[][]{Ananna+19}, the first one closely following the analytic formula by \citetalias[][]{Ueda+14} with updated parameters, and the second one derived from Machine Learning algorithms, labelled as \citetalias[][]{Ananna+19}-ML throughout. {\citetalias{Ananna+19} compiled a large AGN sample with luminosities $\log L_{\rm X}/{\rm erg \ s^{-1}} = 40 - 46$ in the redshift range $0.04 < z < 3.5$, and similarly \citetalias{Buchner+15B} covered luminosity and redshift ranges of $\log L_{\rm X}/{\rm erg \ s^{-1}} = 42 - 46$ and $0 < z < 6$, and \citetalias{Ueda+14} ranges of $\log L_{\rm X}/{\rm erg \ s^{-1}} = 41 - 46$ and $0 < z < 5$. We refer the reader to \citetalias{Alonso-Tetilla}, for a more in-depth overview and comparison of these observational samples}. {It is interesting to note that the cumulative CTK fractions inferred by \citetalias{Ueda+14} are consistent with those recently measured at low redshifts from NuSTAR by \citet[][see also \citealt{Zhang25}]{Boorman25}.}
{\citetalias{Ananna+19}-ML replace the absorption function from \citetalias{Ueda+14} with the one from \citet{Ricci+15}, which includes corrections for the geometry of the torus and inherently predicts a higher fraction of CTK AGN sources, particularly in the local Universe. \citetalias{Ananna+19} suggest that the neural networks allow for a more in-depth exploration of the fraction of CTK sources that simultaneously contribute to the X-ray LF and X-ray background, which could be both underrepresented due to observational biases. To allow a direct comparison of our models with \textit{all} the published data sets considered in this work, we will define CTN fractions as the number of AGN with $22 < \log N_{\rm H}/{\rm cm^{-2}} < 24$  within the larger sample of AGN with $20 < \log N_{\rm H}/{\rm cm^{-2}} < 24$, and CTK fractions as the number of AGN with $24 < \log N_{\rm H}/{\rm cm^{-2}} < 26$ within the full population of AGN with $20 < \log N_{\rm H}/{\rm cm^{-2}} < 26$. For our main reference model, we will also discuss in Section \ref{subsec:AbsoluteNumbersAGN} the predicted absolute number densities of AGN in different intervals of $N_{\rm H}$. } 

\subsection{Comparison strategy: designing models to reproduce high obscured AGN fractions at cosmic noon}\label{sec:comparisonstrategy}

{As discussed in Section~\ref{sec:data}, the observational data considered in this work do \textit{not} provide a clear consensus on the overall fractions of obscured AGN at any redshift, and also on their dependence on AGN luminosity. Some groups suggest a sharp drop in the fractions of obscured AGN with increasing luminosity, such as \citetalias{Ueda+14} and \citetalias{Ananna+19}, a trend which is also supported by optical/UV AGN selections \citep[e.g.,][]{Merloni+14}, while others suggest a mild luminosity dependence, such as \citetalias{Buchner+15B}. In addition, very different results have been retrieved from the same data when using different approaches. \citetalias{Ananna+19}, when adopting machine learning, claimed a nearly constant fraction of at least 80\% of obscured AGN independent of luminosity and redshift. 
Given the noticeable uncertainties and systematics in current observational measurements, in what follows we will retain for completeness all the available data at any given redshift of interest, and discuss the general features a model should possess to 1) induce a pronounced luminosity dependence, as observed in some data sets, and/or to 2) generate an increased incidence of CTN/CTK AGN as suggested by other data sets. In other words, in this work we are not aiming to provide a single preferred model, but rather to gauge the possible classes of physically motivated Orientation and/or Evolution models that could account for different sets of observations. We will show that boosting the overall fractions of obscured AGN, especially in the CTK regime, can be realised with an inner torus, which also helps to generate some luminosity dependence, although we will also discuss some physically plausible alternatives.}

{In this work we focus on the redshift range $1<z<3$, around the peaks of AGN and star formation rate activities \citep[e.g.,][]{Ueda+03,Shankar+09,Madau14}, which are believed to be contributed by galaxies with frequent episodes of dust-enshrouded AGN activity coupled to moderate to strong starbursts, possibly triggered by galaxy mergers \citep[e.g.,][]{Alexander+05,Granato+06,Hopkins+08,AlexanderHickox12,Carraro+20}. This cosmic epoch is thus ideally suited to probe the relative roles of transitory (only for a fraction of the AGN LC) or prolonged (as in Orientation models) obscured phases, induced by the physical characteristics of the host galaxy (as emphasized by, e.g., \citealt{Lapi+14,Gilli+22,Silverman+23,Andonie+23}), and of an inner thick torus, in shaping the fractions of obscured AGN. We verified that over the full redshift range $1<z<3$ the GAEA model does in fact predict that the vast majority of AGN are triggered by {minor and major} mergers, with only up to $\sim 10-15\%$ by disc instabilities, and with many AGN characterised by significant cold gas fractions and large star formation rates. We will show the predictions of our models at the reference redshift of $z=2.4$, and will also briefly discuss the (weak) evolution of the model predictions within the whole redshift range $1<z<3$, but we will not further explore the model performance at lower or higher redshifts. As discussed above, at $z<1$, the radio-mode accretion/feedback starts playing an increasingly relevant role in the overall bolometric emissivity of AGN, adding another channel to the AGN feedback and changing the overall conditions in the host galaxies, whilst at higher redshifts other forms of central obscuration, not necessarily captured in the present modelling, may further contribute to the population of obscured AGN \citep[e.g.,][]{Maiolino+24,BisigelloLRDs, Ma25LRDs}, as briefly mentioned in Section~\ref{subsec:LRDs}.}

\section{Methodology}\label{sec:methodology}

{Starting from the GAEA model galaxy catalogues, {we then assign a degree of obscuration to each AGN as follows}. In pure Evolution models\footnote{{We stress that with the word "Evolution" in this paper we usually refer to the \textit{short} evolution of the SMBH within the timeframe of the AGN light curve and not to the overall evolution of the host galaxy on longer cosmological timescales.}}, we assign a column density by randomly drawing a time within the LC (see Figure \ref{fig:models14}) of each AGN, having a different obscuration pre-peak, post-peak and at the peak. For all our Evolution models, we also study the effect of including a torus-like component. We then define models that combine elements from the Evolution framework and the Orientation scenario (described in \citetalias{Alonso-Tetilla}), {and we label these combinations as {\it hybrid models}}. In these models, as described in \citetalias{Alonso-Tetilla}, we calculate the {HI} column density by defining the geometry of the gas component of the host galaxy using an exponential {gas} density profile, {to which we will also add a torus-like component around the central SMBH}.}  {Finally, for completeness we also explore Evolution models in which the level of obscuration is not determined by the position of the AGN within the LC, but only dictated by the level of starburstiness in the galaxy or by the strength (mass ratio) of the merger.} {A detailed description of all the models considered in this work is provided in the next Sections, while a summary of the main features of each model can be found in Table \ref{tab:models}}.

\begin{table*}
  \centering
  \caption{\textbf{Summary of the different Evolution and Orientation models considered in this work.}}
    \begin{tabular}{|p{8em}|p{9em}|p{37em}|p{3em}|}
    \hline
    \multicolumn{1}{c}{\textbf{Type}} & \multicolumn{1}{|c|}{\textbf{Model}} & \multicolumn{1}{c|}{\textbf{Description}} & \multicolumn{1}{c|}{\textbf{Figure}}  \\
    \hline
    \multirow{4}{8em}{Evolution models}  & Model 1 & CTN pre-peak, visible peak ($\Delta \tau_{\rm QSO} \sim 10^{7-8}$ yr) and post-peak & \multirow{2}{*}{\ref{fig:evolv12}} \\
    & Model 2 & CTN pre-peak, visible peak ($\Delta \tau_{\rm QSO} \sim (1-5)\cdot 10^7$ yr), no post-peak &   \\
    & Model 3 & CTK 1/3 pre-peak, CTN 2/3 pre-peak, visible peak & \multirow{2}{*}{\ref{fig:evolv34}}\\
    & Model 4 & CTK pre-peak, CTN post-peak, visible peak & \\
    \hline
   \multirow{2}{8em}{Evolution + torus} & \multirow{2}{9em}{Models 1-4} & Same LC + torus (long-lived = present throughout the LC; short-lived = present only during obscured parts of the LC) & \ref{fig:empiricaltorus12}, \ref{fig:empiricaltorus34} \\
   \hline
   Evolution + lum dep & Models 1-4 & Same LCs with luminosity dependent visibility window & \ref{fig:evolv12}, \ref{fig:empiricaltorus34} \\
    \hline
    Hybrid model & Models 3-4 & Same LCs + $N_{\rm H}$ from Orientation + long-lived torus & \ref{fig:mixed34} \\
    \hline
    \multirow{1}{8em}{$z$-evolution} & Model 3 & Repeated at different redshifts & \ref{fig:best-fits} \\
    \hline
    \multirow{2}{*}{Starburst} & \multirow{1}{9em}{SFR $>100$ M$_{\odot}$/yr} & $N_{\rm H}$ from Orientation + galaxies with SFR $>100$ M$_{\odot}$/yr are CTK & \multirow{2}{*}{\ref{fig:starburst}} \\
    & \multirow{1}{9em}{SFR $>(1,4) {\rm SFR}_{\rm MS}$} & $N_{\rm H}$ from Orientation + galaxies with SFR $>(1,4) {\rm SFR}_{\rm MS}$ are CTK & \\
    \hline
    Mergers & Merger rate > 0.33 & $N_{\rm H}$ from Orientation + galaxies with merger rate > 0.33 are CTK (w/o fiducial torus) & \ref{fig:mergers}\\
    \hline
    \end{tabular}%
  \label{tab:models}%
\end{table*}%

\subsection{Definition of the Light Curve Model}
\label{subsec:LCs}

Each accretion event onto the SMBH is characterized by a $\Delta M_{\rm BH}$ which is gradually deposited onto the central SMBH following a predefined LC, which regulates the time evolution of the gas accretion rate over the lifetime of the AGN episode. {The total temporal length of the LC is the same for all AGN in the model and equal to $\Delta\tau=2\cdot10^8$ yr (see Figure \ref{fig:SketchBaselineLC}).} The shape of the LC broadly follows expectations from numerical experiments and theoretical arguments \citep[e.g.,][]{Springel+05, Hopkins+06, Lapi+06, Shen09, Shankar10mergermodels}. It is characterized by an exponential pre-peak regime, followed by a post-peak power-law phase, to mimic the effects of a self-regulated growth of the SMBH, which initially grows exponentially at or above the Eddington limit, then switches to a less rapid accretion mode when the gas reservoir reduces \citep[e.g.,][]{Granato+04, Monaco+07, Fontanot+20}. {Both parts of the LC depend on the same set of parameters, as described below.}

The exponential regime is simply described as
\begin{equation}
    \dot{M}_{\rm BH}(t) = \dot{M}_{\rm BH}^{\rm peak} \exp{\left(-\frac{t_{\rm peak}-t}{t_{\rm Edd}} \right)}, \label{eq:prepeak}
\end{equation}
where $\dot{M}_{\rm BH}(t)$ is the accretion rate onto the central SMBH, $\dot{M}_{\rm BH}^{\rm peak}$ is the peak accretion rate in the QSO mode, $t_{\rm peak}$ is the time corresponding to the peak accretion rate since the triggering episode (most of the time a merger), $t$ is the time stamp within the LC, and $t_{\rm Edd}$ is the Eddington time corresponding to $4.5 \cdot 10^7$ years for our chosen value of the radiative efficiency ($\epsilon = 0.15$, also used for luminosity calculation). 

The second regime is defined as in \citet[][]{Hopkins+06},
\begin{equation}
    \dot{M}_{\rm BH} = \frac{\dot{M}_{\rm BH}^{\rm peak}}{1+
    \left|\frac{t-t_{\rm peak}}{t_{\rm Edd}}\right|^2}. \label{eq:postpeak}
\end{equation}

The LC put forward above is a flexible mathematical model that can be adapted to explore the impact of varying the relative time lengths of the pre- and post-peak phases in the predicted fractions of obscured AGN \citepalias[see also Appendix B in][]{Alonso-Tetilla}. 

{{We visualise our baseline LC model in Figure \ref{fig:SketchBaselineLC} which includes two limiting cases of an AGN with a LC with luminosities for most of the time above the minimum bolometric luminosity (mostly from AGN X-ray surveys) $L_{\rm bol}\gtrsim 10^{42}\, {\rm erg s^{-1}}$ (left panel), and one with a LC in which only luminosities around the peak are above $L_{\rm bol}\gtrsim 10^{42}\, {\rm erg s^{-1}}$ (right panel). Most of the sources in our models broadly fall within these two limiting cases, with a larger proportion of AGN with fainter luminosities, as relatively fewer AGN reach very bright luminosities around their peaks.}} 

In the original LC model presented by \citetalias[][]{Fontanot+20} (Eq. 13), the peak values were chosen by adopting a first regime where the BH accretes exponentially at $\dot{M}_{\rm edd}$, until it reaches a critical BH mass $M_{\rm BH}^{\rm crit}$ ($\dot{M}_{\rm BH}^{\rm crit} = \dot{M}_{\rm BH}^{\rm peak}$)

\begin{equation}
    M_{\rm BH}^{\rm crit} = f_{\rm crit}1.07(M_{\rm BH}^{\rm in}+\Delta M_{\rm BH}),\label{eq:mcrit}
\end{equation}
where $\Delta M_{\rm BH}$ represents the total mass accreted in the event, $M_{\rm BH}^{\rm in}$ is the initial mass of the SMBH, and we fix {the scaling factor} $f_{\rm crit} = 0.4$ as in \citetalias[][]{Fontanot+20}, {following} \citet[][see also \citealt{Marulli+08, Bonoli+10}]{Somerville+08}.  

The original \citetalias[][]{Fontanot+20} LC, calibrated on the \citet{Hopkins+07} prescriptions built around the idea of a critical accretion rate $M_{\rm BH}^{\rm crit}$, tends to generate quite narrow LCs, sometimes close to delta functions, due to the fact that after triggering, the accretion rate rapidly approaches the critical accretion rate thus generating a fast switch from Eddington-limited to power-law, sub-Eddington regimes. 

In this work, instead, we {relax the critical accretion rate limitation}. In our LC reference model, we {keep unaltered} the original \citetalias[][]{Fontanot+20} values of $\dot{M}_{\rm BH}^{\rm crit}$ and $t_{\rm peak}$, but allow for extended pre- and post-peak phases following Eqs. \ref{eq:prepeak} and \ref{eq:postpeak}. In addition, we will also explore more flexible models where both the $t_{\rm peak}$ and $\dot{M}_{\rm BH}^{\rm peak}$ parameters are also varied. {We verified that our predicted AGN LFs generated by the new LCs, fall within the observational determinations by \citet{Shen+20} at $z \lesssim 3$ (see Section \ref{sec:discussion}), and also the implied SMBH mass functions are similar to those reported by \citetalias[][]{Fontanot+20}.} 

\subsection{Pure Evolution Models}
\label{sec:empirical}

{In our pure Evolution models, the level of obscuration in an AGN is only controlled by the time at which the AGN is observed within the LC \citep[e.g.,][]{Sanders+88}.} In our Evolution models, each source is randomly sampled within the LC and assigned a luminosity and a certain degree of obscuration (a value of the line-of-sight $N_{\rm H}$ column density) based on its position in the LC. The early phases of the life of an AGN are considered the most obscured, as the AGN radiation is expected to clear out its environment at later times, making the source {also} optically and UV visible \citep[e.g.,][]{Hopkins+06, Gilli+07, Kocevski+15}. However, {several dedicated studies} suggest a more complex evolution \citep[e.g.,][]{Lapi+06, Lapi+14, Aversa+15}. {As visually represented in Figure \ref{fig:models14}, in this work we consider four physically motivated Evolution scenarios that schematically account for the time evolution of the level of obscuration within the timeframe of the AGN LC (without a torus-like component). All four reference Models are based on the same underlying AGN LC presented in \ref{subsec:LCs}, but differ in terms of level of obscuration and/or on the extent of the post-peak phase. The main features of each Model are described below}:
\begin{itemize}
    \item \textit{Model 1}. This model follows the traditional view of Evolution models in which AGN are obscured pre-peak, and become gradually optically/UV visible post-peak. {In Model 1, we assume that the AGN can only be obscured during the pre-peak phase of the LC and reach up to CTN levels, with an assigned random value of the Hydrogen column density within $22<\log N_{\rm H}/{\rm cm^{-2}}<24$}. {In Model 1, the optical/UV ``visibility window'' $\Delta \tau_{\rm QSO}$ is the whole extent of the LC after the time of peak luminosity, labelled as ``unobscured'' in Figure \ref{fig:models14}}.
    \item \textit{Model 2}. {This model is identical to Model 1 but lacks the post-peak phase. This model is inspired by the seminal works by \citet{Granato+04} and \citet{Lapi+06}, which suggest that the source is first obscured until the SMBH feedback and the strong early episodes of star formation efficiently eject and/or consume the gas reservoir in the host galaxy and around the SMBH, thus rapidly shutting off the accretion onto the central object, consequently inducing a sharp decreases of the LC to zero after the peak}. The AGN in this scenario becomes optically/UV visible for a relatively brief interval of time $\Delta \tau_{\rm QSO}$ around the peak. {As in Model 1, we here also assume that the AGN only reaches CTN levels during the obscured phase}. 
    \item \textit{Model 3}. This model is a more {comprehensive} variation of Model 2 {and it can be considered our reference or preferred Evolution model} in which we {also} include an initial Compton-thick (CTK) phase followed by a CTN phase. The peak and post-peak phases are unobscured ($20<\log N_{\rm H}/{\rm cm^{-2}}<22$). This model can adopt a flexible post-peak phase. If we make use of the total LC, we {label} it \textit{with post-peak}. If, similarly to Model 2, the accretion turns off quickly after the peak, we {label the model} \textit{without post-peak}. {In Model 3 the optical/UV visibility window $\Delta \tau_{\rm QSO}$ is the total extent of the LC after the CTN phase}. {We assume that the CTK and CTN phases last, respectively, $1/3$ and $2/3$ of the whole pre-peak duration of the LC, to ensure that at least 30\% of CTK AGN are generated, in line with the measurements by \citetalias{Ueda+14} and \citetalias{Ananna+19}, as further detailed below.}. 
     \item \textit{Model 4}. In this model, the early phases of the AGN present CTK obscuration, followed by a period of $\Delta \tau_{\rm QSO}$ {of} optical/UV visibility. {We then assume that the AGN turns back to a CTN phase at a later, post-peak stage, mimicking the effect of ejected gas falling back onto the galaxy causing a delayed CTN obscuration phase.} \footnote{{In principle, we could assume the CTN phase to occur just after the CTK one around the peak, but this model would generate too few luminous, blue quasars and also, we checked, it does not predict the correct distribution of CTN AGN.}}
\end{itemize}

{The first four rows of Figure~\ref{fig:models14} sketch the four models introduced above. In the left column we report the idealised LC characterizing each model, and in the right column the corresponding foreseen effects on the obscured fractions of AGN as a function of luminosity $f_{\rm obs}(L)$, which will be discussed in detail in Section~\ref{sec:results}. It can be seen from Figure~\ref{fig:models14} that the removal of a prolonged post-peak phase (blue lines) is expected to significantly increase the fractions of obscured AGN $f_{\rm obs}(L)$ at all luminosities as the visibility window $\Delta \tau_{\rm QSO}$ shrinks while the obscured pre-peak phases remain unaltered.  
}

It is important to stress at this point that the overall extent of the AGN LC is fixed in our models to {{$\Delta\tau\sim 2\cdot 10^8$}} years {{(Figure \ref{fig:SketchBaselineLC})}} as in \citetalias{Fontanot+20} at $z=2.4$, although it could be shorter than this in models without a post-peak phase, as, for example, in Model 2. We note that the exact value of the full temporal length of the LC has a relatively minor impact on both the predicted bolometric LF and the obscured fractions of AGN. What is more relevant is the choice of the \textit{relative} lengths of the pre- and post-peak phases, which correspond to different levels of obscuration in our Evolution models. 
In addition, we also verified that most of our sources are already shining above the limiting luminosity of the observational surveys, $L_{\rm bol}\gtrsim 10^{42}\, {\rm erg s^{-1}}$, and thus prolonging the LCs would mostly add luminosities below the detection limit of the data, without any impact on the predicted $f_{\rm obs}(L)$ above $L_{\rm bol}\gtrsim 10^{42}\, {\rm erg s^{-1}}$ {{(see Figure \ref{fig:SketchBaselineLC})}}. 

{Once a model with its LC is chosen, we assign to each AGN a Hydrogen column density $N_{\rm H}$ and a bolometric luminosity $L_{\rm bol}$ by randomly extracting the time of observation of the AGN from its LC}. If we observe the galaxy in a phase of CTN obscuration, we assign $N_{\rm H}$ at random in the range $22<\log (N_{\rm H}/$cm$^{-2})<24$. The optically/UV visible phase is characterized by $20<\log (N_{\rm H}/$cm$^{-2})<22$, while the CTK column densities are uniformly extracted in the range $24<\log (N_{\rm H}/$cm$^{-2})<26$. {In this approach, we bypass any information on the amount and/or geometry of the cold gas within the host galaxy, with the aim to characterize the general conditions under which a pure Evolution sequence of obscured/unobscured phases can reproduce current data sets.} 

\subsubsection{Including a fine-tuned luminosity dependence}
\label{subsec:LuminosityDependence}

{In all our reference models, we generally assume that the optical/UV visibility window $\Delta \tau_{\rm QSO}$ is constant for all galaxies. However, it may be expected that in more luminous AGN, with allegedly an increased ejective power, $\Delta \tau_{\rm QSO}$ may be longer than for less luminous sources. This process would, in turn, generate shorter obscuration phases, and thus a reduced fraction of obscured sources at higher luminosities, as visually sketched in the first rows of Figure~\ref{fig:models14} with orange, dashed lines}. In order to test this possibility, we assume $\Delta \tau_{\rm QSO}$ to slightly increase with peak luminosity in {{Models 1 and 3}}, following the empirical formula 
\begin{equation}
    \Delta \tau_{\rm QSO}(L_{\rm peak}) = 
         \Delta t_{\rm post-peak}\cdot \left[1 - \left(\frac{L_{\rm lim}}{L_{\rm peak}}\right)^{\alpha} \right] \cdot 10^7 \rm \left[ yr \right], \label{eq:lumdep}
\end{equation}
where $\alpha = 0.1$, $L_{\rm lim} = 10^{45}$ erg/s, and $\Delta t_{\rm post-peak} = f(t_{\rm peak})$ refers to the time of the post-peak phase which varies from source to source, but limited in the range $10^7$ and $2$ $\cdot 10^8$ years, the maximum extent of the LC at our redshift of reference $z=2.4$. We also impose that $\Delta \tau_{\rm QSO}$ never falls below $10^7$ yr, i.e., the reference value for models with constant visibility window, as we noticed that this choice provides a better match to the fraction of obscured AGN at fainter luminosities. Similarly, {{Models 2 and 4 have}} a fine-tuned visibility window:
\begin{equation}
    \Delta \tau_{\rm QSO}(L_{\rm peak}) = 
         \Delta t_{\rm pre-peak}\cdot \left[1 - \left(\frac{L_{\rm lim}}{L_{\rm peak}}\right)^{\alpha} \right] \cdot 10^7 \rm \left[ yr \right].  \label{eq:lumdep2}
\end{equation}

Note that in {{Models 1 and 3}} we are fine-tuning the visibility window by decreasing the time of the post-peak phase {(Eq. \ref{eq:lumdep})} and in {{Models 2 and 4}} we use the pre-peak phase instead {(Eq. \ref{eq:lumdep2})}. Therefore, in {Models 1 and 3 we keep the CTN/CTK phases constant, while in Models 2 and 4 we decrease them} as the optical/UV visibility window increases. {The form and choice of parameters for the luminosity-dependent expressions of $\Delta \tau_{\rm QSO}$ given above have been empirically calibrated via trial and error to produce an improved match to the fractions of obscured AGN measured by \citetalias{Ueda+14} and \citetalias{Ananna+19}, which sharply drop with increasing AGN luminosity. Equations \ref{eq:lumdep} and \ref{eq:lumdep2} ensure that the more luminous AGN will be characterized by longer visibility windows, more specifically AGN with peak luminosity $L_{\rm peak} \lesssim 10^{43}$ erg/s will have a $\Delta \tau_{\rm QSO} \sim 10^7$ yr, which steadily increases for more luminous sources approaching values of $\Delta \tau_{\rm QSO}\sim 5\cdot 10^7$ yr for $L_{\rm peak} \lesssim 10^{45}$ erg/s and $\Delta \tau_{\rm QSO}\sim 8\cdot 10^7$ yr for $L_{\rm peak}\sim 10^{46}$ erg/s.}  
 
\subsubsection{Including a central torus-like component}
\label{subsec:IncludingTorus}

The Evolution models introduced so far do not include a possible contribution from a central torus-like structure. However, it is now clear from both direct and indirect (via, e.g., Spectral Energy Distribution (SED) fitting) observations \citep[][]{Combes+19, Garcia-Burillo+19, Garcia-Burillo+21} that such element is an essential ingredient required to fully model the observational properties of AGN \citep[see][for reviews]{Netzer15, Ramos-Almeida+17, Hickox+18}, especially in sources with $\log_{10} (N_{\rm H}/{\rm cm}^{-2}) > 24$ \citep[][]{Risaliti+99, Marchesi+18}. As already mentioned in \citetalias{Alonso-Tetilla}, the torus can be described as a compact reservoir of low angular momentum dusty gaseous material, and/or part of a windy outflowing structure connected to the accretion disc \citep[][and references therein]{Hoenig19}. Irrespective of its underlying nature, a torus around a SMBH significantly contributes to absorb UV light from the accretion disc and reprocess it in IR bands. Despite GAEA includes the modelling of gas reservoirs around the central SMBH (and the subsequent accretion of this material), it does not explicitly treat the dynamical and geometrical properties of the accretion disc and the torus around the central SMBH. Therefore, we include a modelling of the torus following the prescriptions we developed in \citetalias{Alonso-Tetilla}. We define as our \textit{fiducial} torus model a combination of the models proposed by \citet[][]{Wada15} \footnote{We adopt the version of the code provided by Johannes Buchner: \url{https://github.com/JohannesBuchner/agnviz}} and \citet[][]{Ramos-Almeida+17}. The former model analytically connects the dependence of the torus size and thickness on AGN luminosity/accretion rate and SMBH mass, and assumes that in an AGN there is always enough circumnuclear material to feed a torus. {The latter more empirical model by \citet[][]{Ramos-Almeida+17} assumes that the column density increases for larger inclination angles, with maximum CTK column densities for lines of sight close the centre of the torus, without any explicit dependence on SMBH accretion rate or mass (see \citetalias{Alonso-Tetilla} for full details)}. 

{In what follows, we will explore variants of our reference Evolution Models 3 and 4 in which we add the torus component, as sketched in the bottom row of Figure~\ref{fig:models14}. We label as \textit{short-lived} those Evolution models where we assume the torus is only present during the obscured phases within the LC, while we label \textit{long-lived} those Evolution models where we assume the torus survives the peak activity of the AGN and lasts for the whole duration of the LC. Including a torus will always boost the fraction of obscured AGN, especially in the CTK regime, and will also tend to produce obscured fractions that decrease with increasing luminosity as the thickness of the torus itself shrinks with AGN luminosity \citep[][]{Wada15}. Therefore, the torus model is expected to provide predictions that may be somewhat degenerate with those from a luminosity-dependent $\Delta \tau_{\rm QSO}$ (Section~\ref{subsec:LuminosityDependence}), and we will discuss some implications of this in Section \ref{sec:discussion}}.  

\subsection{Orientation and Evolution \textit{hybrid} models}\label{subsec:hybridmodels}

{Including a torus component in an Evolution model, as described in the previous Section, is a first step towards a more complete model for the obscuration of AGN. In addition, a more realistic modelling of AGN obscuration should take into account the overall distribution of cold gas in the host galaxy during the LC, e.g., during the growth episode of the central SMBH. In the real Universe, we would thus expect that a combination of Orientation and Evolution effects could simultaneously contribute to the line-of-sight $N_{\rm H}$ column density of an AGN \citep[e.g.,][]{Hickox+18, Zhou+18, Gilli+22, Pouliasis+24}.} To this purpose, in this work we also put forward comprehensive models in which the $N_{\rm H}$ column densities originate from both the gas distribution in the host galaxy and the central torus.  

{Adopting the exponential gas density profile described in \citetalias{Alonso-Tetilla}}, we then assign a bolometric AGN luminosity to the SMBH based on the retrieved value of $N_{\rm H}$ and the type of LC assumed in the model{, following the different LC models described in Section \ref{sec:empirical}}. For example, {using} Model 1, {a density profile-based} column density of $\log_{10} (N_{\rm H}/{\rm cm^{-2}}) < 22$ would have a bolometric luminosity chosen at random within the visible post-peak phase of the LC, whilst column densities $\log_{10} (N_{\rm H}/{\rm cm^{-2}}) > 22$ would have assigned a random bolometric luminosity within the obscured pre-peak phase of the LC. {An equivalent approach is taken for Models 3 and 4, for which we take into account the three portions of the LC, including the CTK one. We label this more comprehensive version of our models \textit{hybrid} since they includes key elements from both the Orientation (density profile gas distributions and central torus) and Evolution frameworks (AGN light curves).}

{As already mentioned above, in the Orientation framework the} column density $N_{\rm H}$ associated to each galaxy is computed following the modelling of \citetalias{Alonso-Tetilla}. In brief, this model strictly assumes an exponential profile for the gas component, with a scale radius of $R_{\rm d, gas} = 0.3\cdot R_{\rm d, \star}$, which aligns with present ALMA and JWST observations of high-z galaxies \citep[see, e.g.][]{Puglisi+19,Price25}. {We will discuss in Section \ref{subsec:differentgasprofile} the implications of relaxing the assumption of a strictly exponential gas density profile.}

\subsection{High star-forming galaxies and major mergers as CTK sources}
\label{subsec:StarburstsMergerModels}

In the context of Evolution models \citep[e.g.,][]{Granato+04, Alexander+05, Granato+06, Hopkins+10}, it is expected that newly formed, dust-enshrouded galaxies, often characterized by intense starburst episodes possibly triggered by major mergers, and usually associated with high gas column densities \citep{Mihos+96, DiMatteo+08, Zhou+18, Renaud+22}, are the obvious sites of obscured CTK AGN, especially at $z>1$. It is thus a natural question to ask whether there are sufficient starbursts or major mergers to explain the significant fractions of CTK AGN observed at different redshifts.

{To this purpose, in this work we will also briefly explore models where all obscured CTK AGN are simply defined to be those galaxies in GAEA characterized by a star formation rate (SFR) above some ad-hoc threshold above the Main Sequence}. In other words, in this alternative approach we are not assuming any underlying geometry for the cold gas, but simply relying on the level of SFR to label an AGN as CTK or not, while the CTN AGN continue to be simply defined by the intrinsic gas content and geometry of the host galaxy. The main limitation to this approach is to use a suitable definition for starburst galaxies. In the following, we take advantage of the GAEA model and label as CTK sources those that are selected above a threshold at their SFR peak. Alternatively, we also adopt a different definition of starburst galaxies as those objects lying four times above the main sequence, which is a common definition of starburst often adopted in observational studies \citep[e.g.,][]{Carraro+20}.

\begin{figure*}
    \centering
    \includegraphics[width=0.9\textwidth]{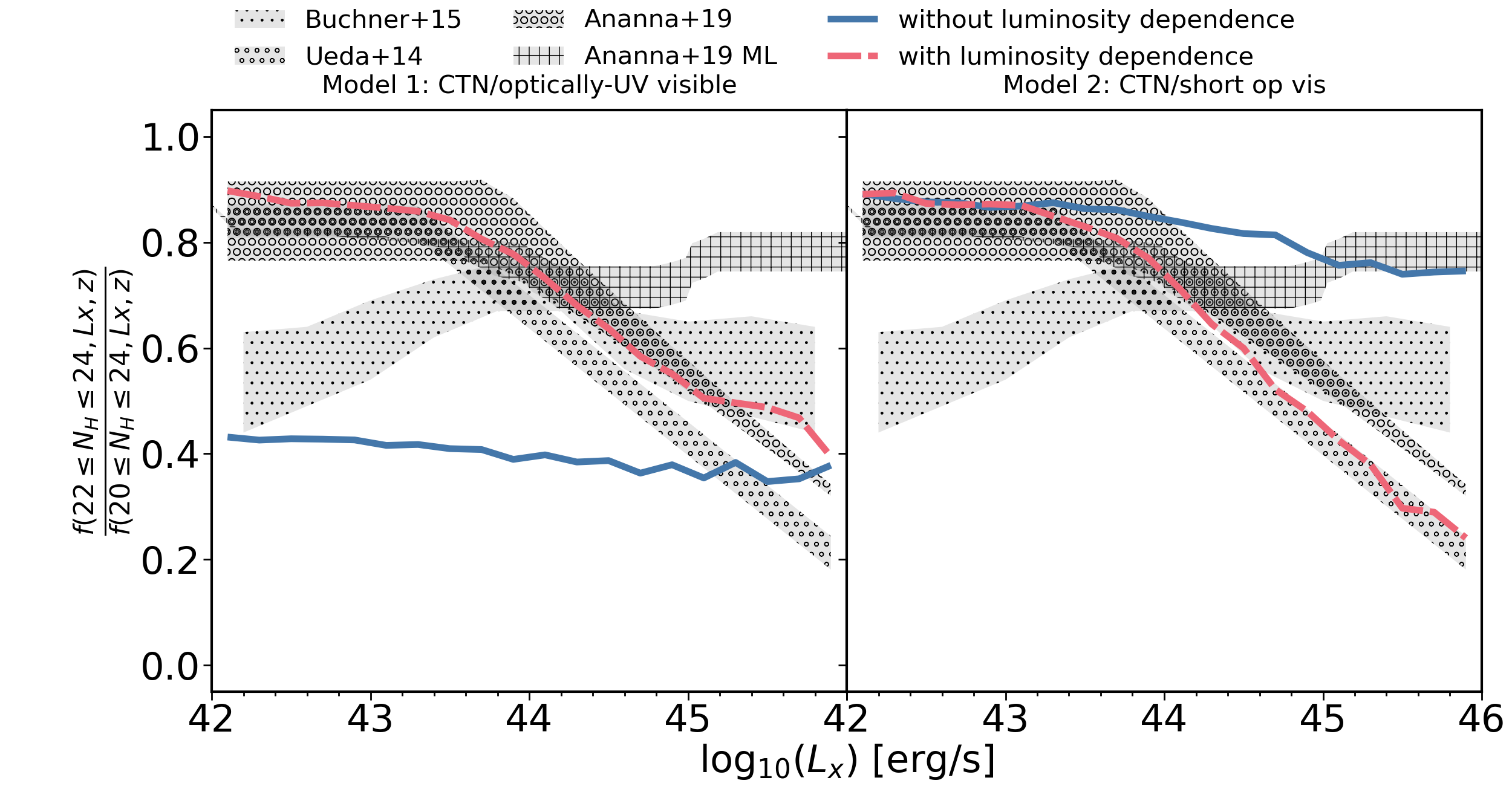}    
    \caption{{Fractions of obscured CTN AGN as a function of X-ray luminosity predicted by Models 1 and 2. The Hydrogen column density is assigned to each host galaxy based on the pre-defined light curve in each Model (see Figure 1). The long-dashed red lines refer to the model variants with luminosity-dependent visibility window $\Delta \tau_{\rm QSO}(L_{\rm peak})$, and the blue solid lines show the predictions of the same models with a fine-tuned luminosity dependency (see text for more details). The observations correspond with \citetalias{Ueda+14}, \citetalias{Ananna+19} and \citetalias{Buchner+15B}, as labelled. Models with luminosity dependent visibility windows better align with the \citetalias{Ueda+14, Ananna+19} CTN fractions. In all Figure labels ``$N_H$'' indicate the logarithm of the Hydrogen column density.}}
    \label{fig:evolv12}
\end{figure*}

Similarly, we can assume that CTK AGN are mainly connected with merger events \citep[e.g.,][]{Lanzuisi+15, Bickley+24}. GAEA allows us to track the events responsible for triggering AGN activity (either disc instabilities or mergers). At the redshift of interest, $z=2.4$, the large majority of simulated AGN ($\gtrsim 90\%$) are the result of mergers (either minor or major). We then explore a variant of our obscuration models in which CTK sources are those galaxies that have undergone a recent merger above a chosen mass ratio.


\section{Results}\label{sec:results}

In this Section we present the main results of assuming different LCs as inputs in the GAEA model, as detailed in Section \ref{sec:methodology}. We will start showing the predictions from Models 1 and 2, and then move to those from Models 3 and 4, in Section \ref{subsec:purevolution}. The effect of including a {central, torus-like component} is shown in Section \ref{sec:evolvwithtorus}, whilst the outputs of the Orientation and Evolution hybrid model are given in Section \ref{sec:mixedmodel}, and the predictions of the reference models at different redshifts in Section \ref{sec:redshift}. Finally, the results from alternative models based on SFR and merger thresholds are presented in Sections \ref{sec:starburst} and \ref{sec:mergers} respectively.

\subsection{Predicted fractions of obscured AGN from pure Evolution models} \label{subsec:purevolution}

\begin{figure*}
    \centering
    \includegraphics[width=0.9\textwidth]{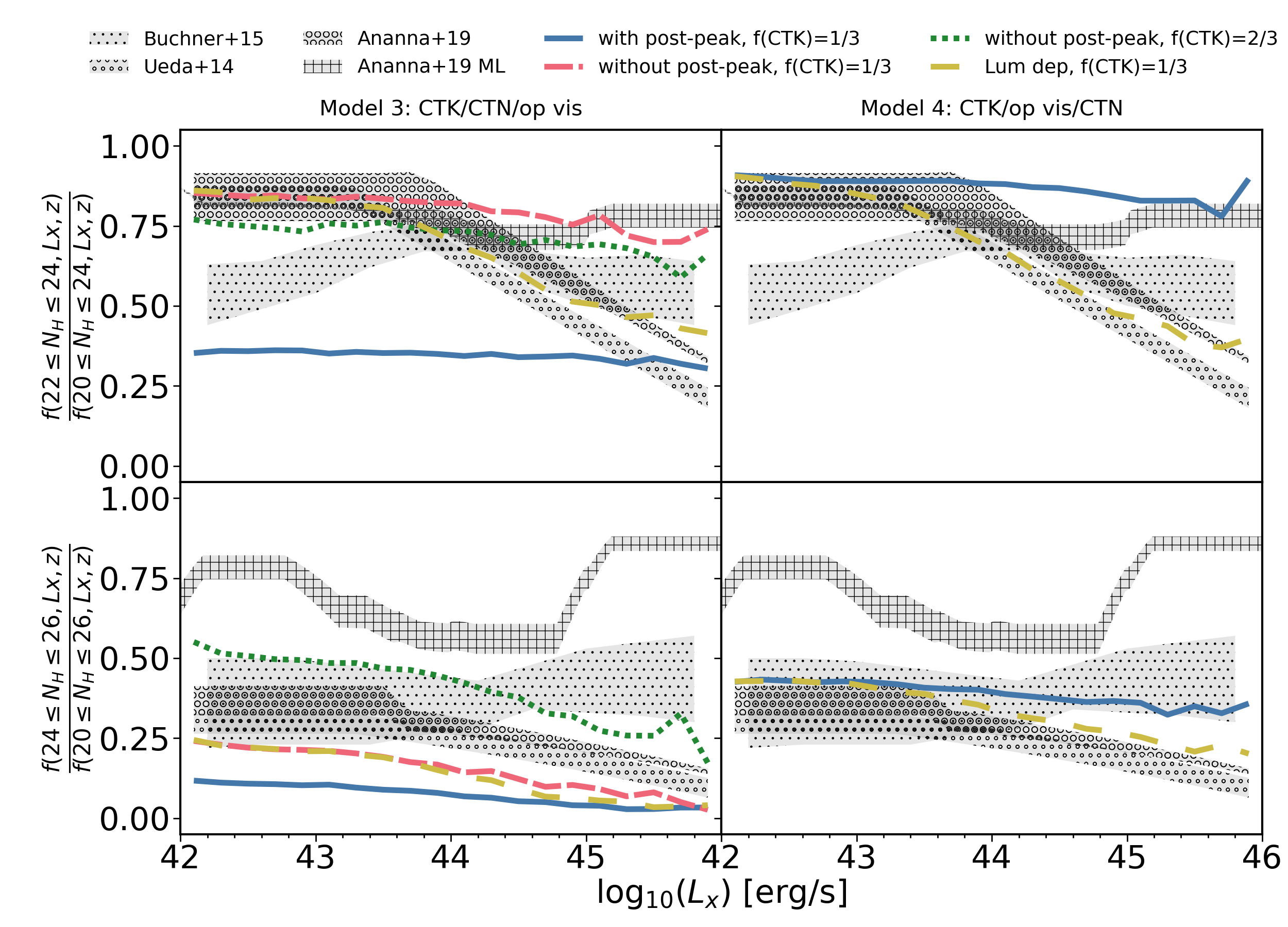}    
    \caption{{CTN and CTK obscured fractions of AGN as a function of X-ray luminosity predicted by Models 3 and 4. Solid blue, long-dashed red, and long-dashed yellow lines refer to variants of the Models with a post-peak phase,  without a post-peak phase, and with a luminosity dependent $\Delta \tau_{\rm QSO}(L_{\rm peak})$ window, respectively. Also shown are a Model 3 variant with a double pre-peak CTK phase (dotted green line). All the data are as in Figure \ref{fig:evolv12}. $f(CTK)$ only affect Model 3, as Model 4 assumes as CTK all the pre-peak phase. A luminosity dependent visibility window model with a post-peak LC phase that decreases with luminosity improves the alignment of Models 3 and 4 to  the \citetalias{Ueda+14, Ananna+19} CTN fractions. Removing the visible post-peak phase in Model 3 increases the fractions of obscured AGN, as expected}.}
    \label{fig:evolv34}
\end{figure*}

\begin{figure*}
    \centering
    \includegraphics[width=0.9\textwidth]{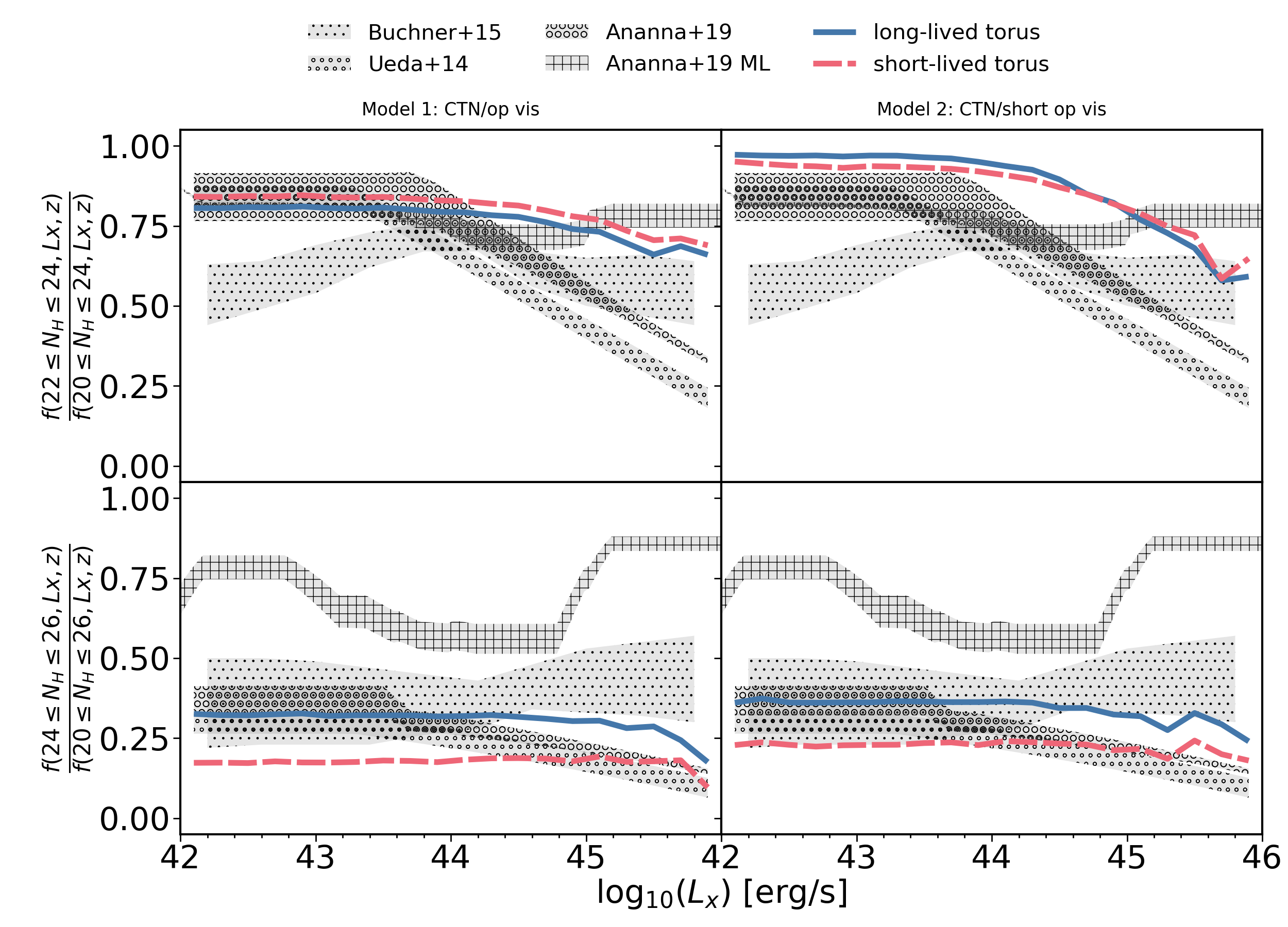} 
    \caption{{CTN and CTK obscured fractions of AGN as a function of X-ray luminosity predicted by Models 1 and 2 inclusive of a torus. The torus can be long-lived (surviving throughout the LC; solid blue lines) or short-lived (present only during the obscured phases of the LC; dashed red lines). The data are as in Figure \ref{fig:evolv12}. A torus component tends to increase the fractions of CTK AGN}.}
    \label{fig:empiricaltorus12}
\end{figure*}

\begin{figure*}
    \centering
    \includegraphics[width=0.9\textwidth]{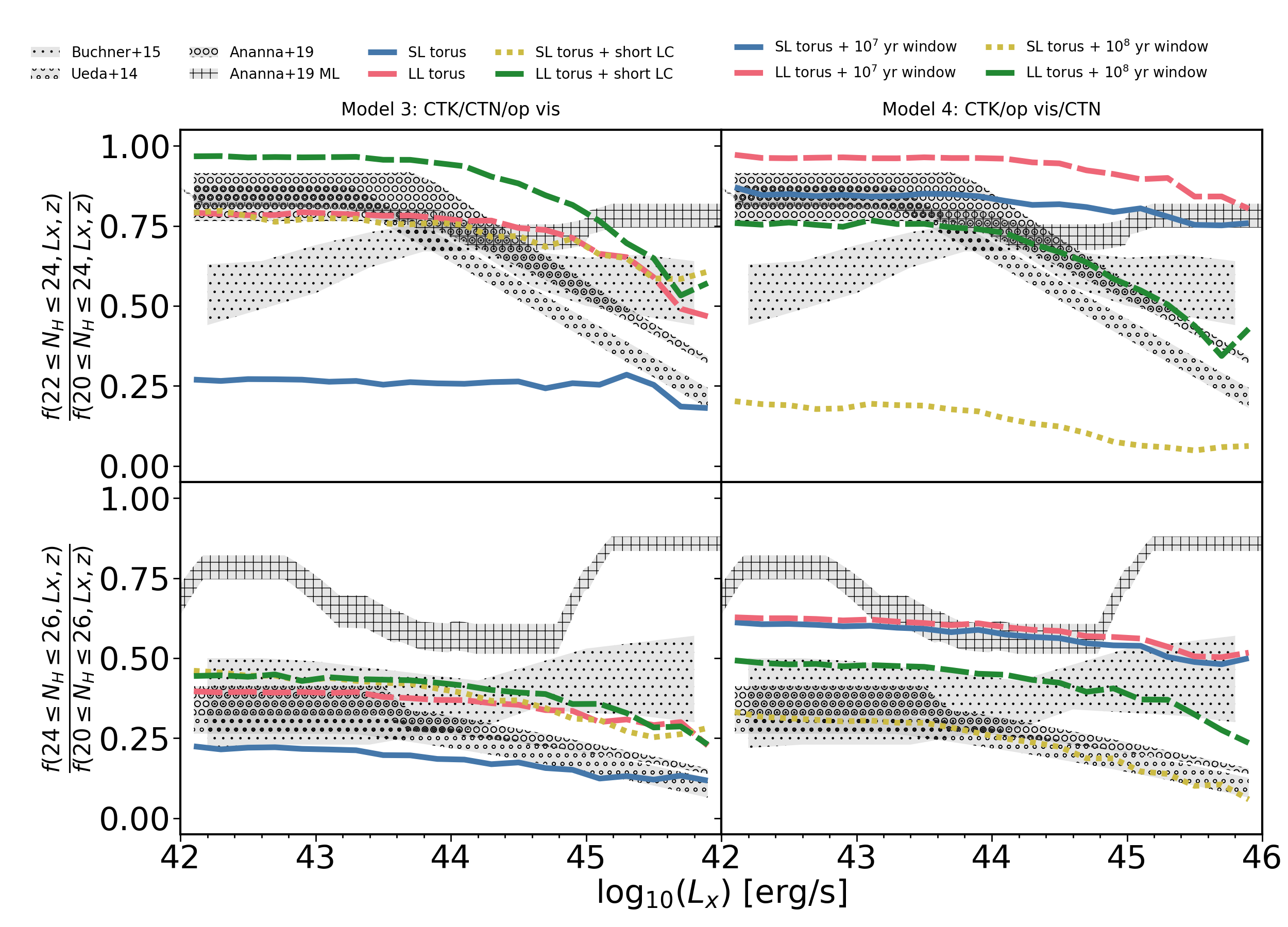} 
    \caption{{CTN and CTK obscured fractions of AGN as a function of X-ray luminosity predicted by Models 3 and 4 with a torus component. Solid blue and yellow dashed lines refer to model variants with a short-lived torus, while dashed red and green lines refer to a long-lived torus. In Model 3 (left panel) the yellow dotted and green dashed lines refer to models without a post-peak phase (``short LC''), while the same lines in Model 4 (right panel) refer to models with a longer optical/UV visibility window $\Delta \tau_{\rm QSO} \sim 10^8$ yr. The inclusion of a long-lived torus boosts the fractions of both CTN and CTK AGN, but this becomes more evident in model variants with a prolonged, post-peak phase and/or a longer visibility window, which would otherwise imply larger fractions of unobscured AGN}.}
    \label{fig:empiricaltorus34}
\end{figure*}


We show the CTN fractions depending on the X-ray luminosity of the first two Evolution models in Figure \ref{fig:evolv12}. In Model 1 (left panel), the obscuration is present in the pre-peak luminosity phase followed by an optically/UV visible post-peak phase. In Model 2 (right panel) we assume that the LC goes rapidly to zero right after the peak. The AGN becomes, therefore, optically/UV visible only at the peak luminosity, being CTN until then. In this case, we find a higher fraction of CTN galaxies compared to Model 1, as expected since we are decreasing the probability of a galaxy to be optically/UV visible with respect to Model 1 which has an extended post-peak phase. These models by design do not produce any CTK obscuration.

{A constant visibility window $\Delta \tau_{\rm QSO}$ (blue solid lines in Figure \ref{fig:evolv12}) tends to produce a rather flat fraction of CTN AGN as a function of luminosity in both Models 1 and 2, with the latter Model generating large CTN fractions in line with what derived from \citetalias{Ananna+19}-ML. In Model 1 the \(\Delta \tau_{\rm QSO}\) encompasses nearly the entire post-peak LC, with more than 50\% of the AGN having \(\Delta \tau_{\rm QSO} \sim 2 \times 10^{8}\) yr, limiting the fraction of CTN AGN to $\sim 40\%$. Conversely, Model 2 is by design characterized by a shorter constant visibility window \(\Delta \tau_{\rm QSO} \sim 10^7\) yr for all AGN, which induces a significantly larger fraction of CTN AGN $\sim 80-90\%$.}

{We also show in Figure \ref{fig:evolv12} the effects of assuming a luminosity dependence in $\Delta \tau_{\rm QSO}(L_{\rm peak})$ (long-dashed red lines), following the empirical expressions given in Eqs. \ref{eq:lumdep} and \ref{eq:lumdep2}. With a luminosity dependent $\Delta \tau_{\rm QSO}(L_{\rm peak})$, both Models 1 and 2 can produce a good match to the \citetalias{Ueda+14, Ananna+19} data, with roughly similar visible timescales, ranging from \(\Delta \tau_{\rm QSO} \sim 10^7\) yr at $\log L_X~\sim 43$ erg/s, to \(\Delta \tau_{\rm QSO} \sim 10^8\) yr at $\log L_X \sim 46$ erg/s, implying a steady decrease in the fraction of obscured AGN toward brighter luminosities. We note that the inclusion of a luminosity dependence in Model 1 boosts the fraction of obscured CTN AGN especially at fainter luminosities where the \(\Delta \tau_{\rm QSO}\) visibility windows shrink up to an order of magnitude for many sources.}. 

In Figure \ref{fig:evolv34} we show the predictions for Models 3 and 4 (left and right panels, respectively) for both CTN and CTK AGN fractions (top and bottom panels, respectively), {since Models 3 and 4 also include CTK obscuration}. Model 3 has two variants, one without post-peak phase \citep[inspired by the results of, e.g.,][]{Lapi+06, Shankar10mergermodels, Aversa+15, Lapi+14} and one with a total post-peak phase \citep[e.g.,][]{Shen+20}. In the former, we assume the AGN in the initial 1/3 of the LC to be in a CTK phase, followed by a longer CTN phase ($\sim$2/3 of the LC) and a constantly short visible window of $\Delta \tau_{\rm QSO}=10^7$ yr. {Our choice of reserving $1/3$ and $2/3$ of the LC in Model 3 to, respectively, the CTK and CTN phases, is simply to broadly align the model predictions with the observed fractions of CTK/CTN AGN. The rendition of Model 3 without post-peak phase (green dotted line), predicts noticeable fractions of CTK AGN up to $\sim50\%$, and up to $\sim75\%$ of CTN AGN, with relatively weak evolution with AGN luminosity, especially for CTN AGN. By decreasing the CTK phase to $1/3$ of the pre-peak phase LC, proportionally decreases the CTK fraction and only slightly increases the CTN one (dashed red lines). Thus, assuming a fraction of CTK AGN of the pre-peak LC to be within $1/3-2/3$, tends to produce sufficient CTK AGN, whilst maintaining a sizeable fraction of CTN AGN, in line with the data. In principle, to increase the fraction of CTK AGN to the level measured by \citetalias[][]{Ananna+19}-ML of up to $\sim 80\%$, would require to proportionally increase the fraction of CTK within the light curve, but inevitably decreasing the fraction of CTN AGN.} Including the post-peak phase in Model 3 (solid, blue lines) boosts the unobscured optical/UV phase by largely decreasing the predicted fractions of CTN and CTK AGN. {In this model, $\sim 50\%$ of the AGN present $\Delta \tau_{\rm QSO} \sim 10^8$ yr, and shorter windows for the rest of the galaxies, similarly to Model 1. With an extended post-peak LC it is thus challenging to simultaneously reproduce the large fractions of CTN and CTK fractions in pure Evolution Models, unless we either allow for some portions of the LC to be obscured even within the peak/post-peak phase and/or assume some luminosity dependence in the visibility window, as we showed in Figure \ref{fig:models14}. We discuss both alternative options below}.

Model 4, reported in the right panel of Figure \ref{fig:evolv34}, is characterized by {a fully CTK pre-peak phase and a long post-peak phase, inclusive of an optical/UV visible window around the peak of the LC of $\Delta \tau_{\rm QSO} \sim 10^7$ yr, followed by a prolonged CTN phase (bottom of Figure \ref{fig:models14})}. Model 4 tends to naturally produce a large fraction of up to $45\%$ of CTK sources and a fraction of CTN sources similar to Model 3 without a post-peak phase. {We note that, in principle, in Model 4 the CTN region could also be placed just after the CTK phase, to allow for some temporal progression in the depletion of the gas content, yielding similar results on the relative fractions of obscured AGN. However, this version of Model 4 would imply that AGN appear predominantly optical/UV visible only towards the end of their LCs and not around the peak of their activity, a possibility that could be tested via, e.g., precise clustering data at high redshifts, as prolonged LCs would imply an inevitable decrease of the large-scale bias, if optical and luminous quasars appear only a long time after the triggering of the AGN \citep[e.g.,][]{Hopkins07clustering,Shankar10mergermodels,Aversa+15}.} 

Both Models 3 and 4, irrespective of the exact length of their LCs, tend to predict rather flat fractions of obscured AGN with X-ray luminosity. To induce a more marked luminosity dependence in the predicted AGN fractions we adopt the luminosity dependent $\Delta \tau_{\rm QSO}(L_{\rm peak})$ given in Eqs. \ref{eq:lumdep} and \ref{eq:lumdep2}, respectively. The latter generate a strong luminosity dependence especially in the CTN AGN fractions, as in Figure \ref{fig:evolv12}. The effect is very similar to Models 1 and 2. For Model 3, $\Delta \tau_{\rm QSO}$ is $ 10^7$ yr for most galaxies, having some galaxies with windows up to $8\cdot10^7$ yr for higher luminosities. In Model 4, a similar situation applies, with windows going up to $\sim 1.1 \cdot 10^8$ yr.

All in all, we conclude that reproducing the observed high fractions of CTN/CTK AGN in pure Evolution models, like the ones explored here, requires either LCs without a post-peak phase or LCs with a substantial portion of the peak/post-peak phase still in the CTN regime, a condition that should be tested against detailed AGN feedback models and observations. For example, according to \citet{Menci+2019}, AGN-driven outflows can reach distances of 20 kpc in about $10^7$ yr perpendicular to the disc, but take around $10^8$ yr to reach the same distance within the plane of the disc, implying that gas expulsion along the plane of the disc may be less efficient within a single AGN lifetime. {This effect could contribute to the generation of obscured post-peak phases, where the AGN may eventually reverse back to a CTN regime in conditions of less efficient clearing of the gas along the plane of the disc}.

\subsection{The impact of a torus component within an Evolution model} \label{sec:evolvwithtorus}

Many groups have found clear evidence of the presence of a torus at the centre of local AGN observed with sufficient sensitivity and resolution \citep[see][for molecular and dusty tori with ALMA, and infrared observations]{Garcia-Burillo+16, Combes+19, Garcia-Burillo+19, Garcia-Burillo+21, Gamez-Rosas+22, Isbell+22, Isbell+23, GarciaBurillo24}. In this Section we explore the impact of including a torus component within our Evolution models. We assume that the torus is a short-lived structure that only contributes to the obscuration along the line of sight during the CTK/CTN phases of the AGN, or we can also assume that the torus is a long-lived structure that survives the AGN feedback blowout {beyond the peak luminosity of the LC}. 

In Figure \ref{fig:empiricaltorus12}, we show that including in Evolution Models 1 and 2 (without the luminosity dependency) our reference torus model from \citetalias{Alonso-Tetilla} (see Section 3.1.1), calibrated on the most recent observations in the local Universe, boosts the fractions of obscured CTK sources (bottom panels), along with the CTN ones (top panels), in better agreement with the data. Choosing for the torus to be present only during the CTN/CTK phase (short-lived, long-dashed red lines) or during the whole AGN LC (long-lived, solid blue lines), has a relatively modest effect in these models, only increasing the CTK fractions by $\sim$20\%. The predicted obscured AGN fractions now also tend to show more pronounced dependence on AGN luminosity due to the inclusion of the torus and its {luminosity-dependent} opening angle, while the visibility window $\Delta \tau_{\rm QSO}$ is not affected by the torus. {As discussed in \citetalias{Alonso-Tetilla}, the torus tends to produce fewer obscured sources at high luminosities because, in the \citetalias{Wada15} model, the opening angle of the torus increases with AGN luminosity, thus reducing the solid angle obscured along the line of sight}.

\begin{figure*}
    \centering
    \includegraphics[width=0.9\textwidth]{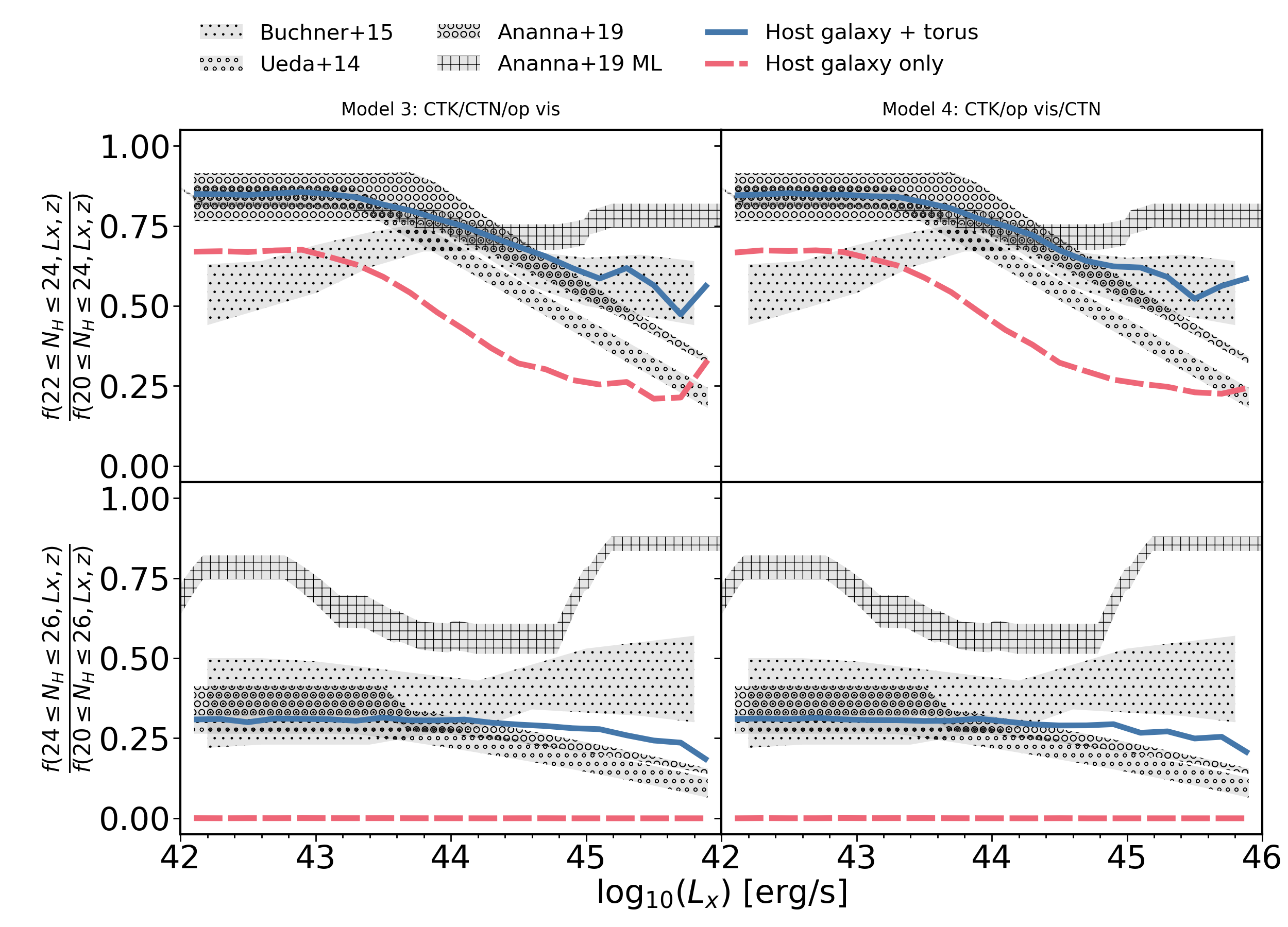}    
    \caption{CTN and CTK fractions of obscured AGN as a function of X-ray luminosity for the \textit{hybrid} Models 3 and 4, inclusive of a Hydrogen column density calculated on an object-by-object basis using the gas content along the line of sight, and a luminosity assigned from the light curve depending on the level of obscuration. In this Figure the torus is always considered long-lived. The optically/UV visible window of Model 4 is $\Delta \tau_{\rm QSO} \sim 10^7$ yr, while Model 3 includes the post-peak phase. The data are as in Figure \ref{fig:evolv12}. The Evolution component has a negligible impact on hybrid models which are mostly regulated by the Orientation elements in the Model. The torus always dominated the number counts of especially CTK AGN.}
    \label{fig:mixed34}
\end{figure*}

In Figure \ref{fig:empiricaltorus34} we show the predictions of Models 3 and 4 with the addition of a long- and short-lived torus. In the left panel of Figure \ref{fig:empiricaltorus34} we report the predicted fractions of CTN/CTK AGN for Model 3 with a post-peak phase and with a long- and short-lived torus component (solid blue and dashed red lines, respectively). It is clear that {in the variants of Model 3} with a post-peak phase LC, the inclusion of a torus only has a significant impact in boosting the CTN/CTK AGN fractions  when it is long-lived, as expected given the relatively short CTK phases. Alternatively, when eliminating the post-peak phase in the LC in Model 3, both the short- and long-lived torus models (dashed green and dotted yellow lines, respectively) tend to boost the CTN/CTK fractions, {because in both cases the torus lives for a significant fraction of the LC. Interestingly, the inclusion of a torus generates a pronounced luminosity dependence in CTN/CTK AGN fractions, especially if long-lived, in line with the \citetalias{Ueda+14, Ananna+19} data, as in Figure \ref{fig:empiricaltorus12}.} 

The right panel of Figure \ref{fig:empiricaltorus34} shows the predictions of Model 4 with the inclusion of a long- and short-lived torus component and a short UV/optical visibility window of $\Delta \tau_{\rm QSO} \sim 10^7$ yr (solid blue and dashed red lines, respectively) and a longer $\Delta \tau_{\rm QSO} \sim 10^8$ yr (dotted yellow and dashed green lines, respectively). In the former case, with a short visibility window, the predicted obscured AGN fractions are similar for both short- and long-lived tori models, as expected given that the difference in the torus lifetime is only $10^7$ yr, in this case. However, when the optical/UV visibility window is increased to $\Delta \tau_{\rm QSO} \sim 10^8$ yr, the difference in the predicted fractions becomes more noticeable (dotted yellow and dashed green lines). {In addition, only one variant of Model 4 presents significant luminosity dependence in line with the \citetalias{Ueda+14, Ananna+19} data, namely the long-lived torus with $\Delta \tau_{\rm QSO} \sim 10^8$ yr (dashed green lines).} 

{All in all, we conclude that the addition of a torus can have profound implications on the predicted CTN/CTK LCs in Evolution models as long as it is assumed to be a long-lived structure around the central SMBH for a significant fraction of the entire LC}.

\begin{figure*}
    \centering
    \includegraphics[width=0.9\textwidth]{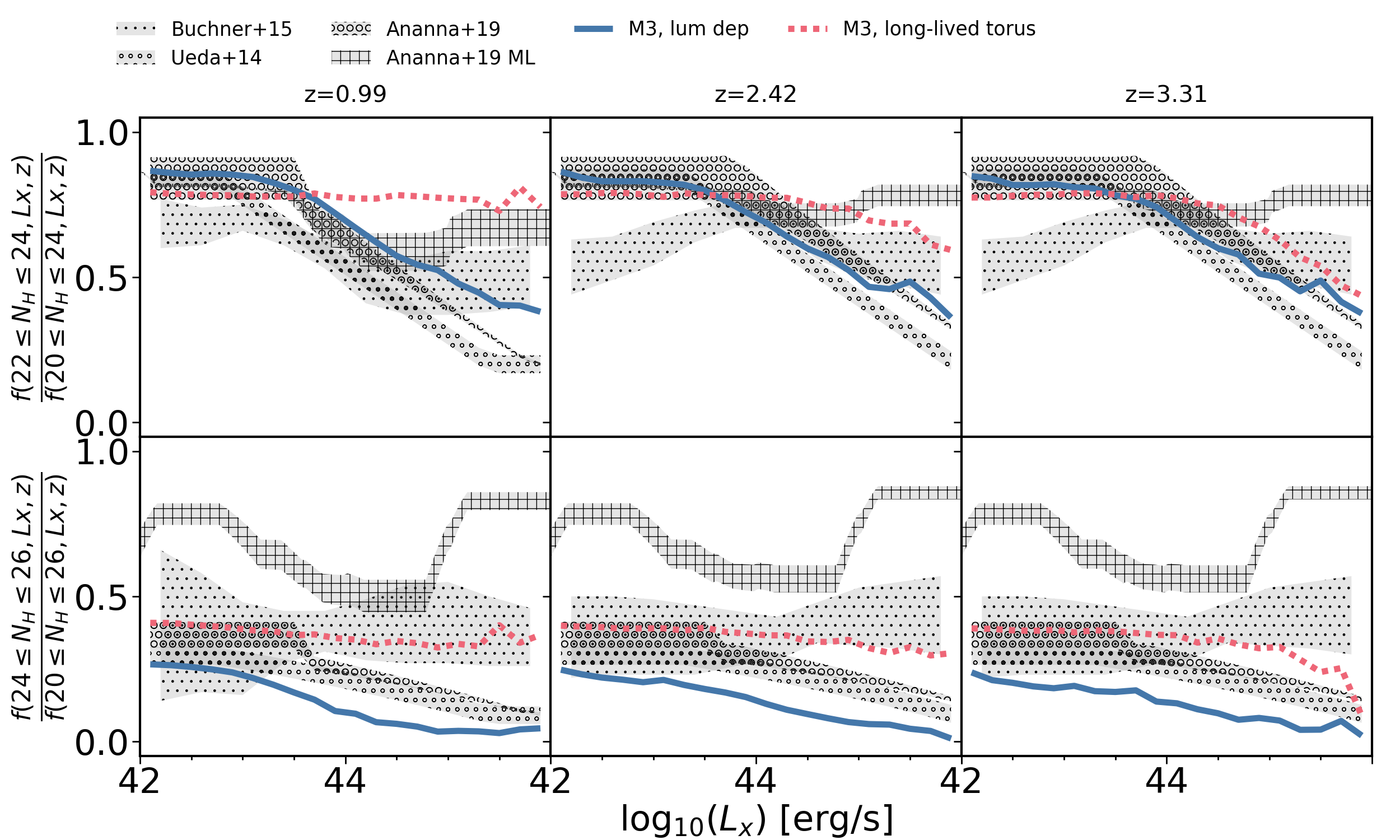}    
    \caption{{Fractions of CTN and CTK AGN predicted by Model 3 with a luminosity dependence visibility window (solid blue lines) and with a long-lived torus (dotted red lines) at redshifts $z=1.0, 2.4,$ and $3.3$. The data are as in Figure \ref{fig:evolv12} at the different redshifts. The predicted fractions are similar at all redshifts, with a slight flattening of the obscured AGN fractions in the model variant with a torus component}.}
    \label{fig:best-fits}
\end{figure*}

\subsection{Hybrid Models: Combining galaxy-based $N_{\rm H}$ column densities and Evolutionary Light Curves} \label{sec:mixedmodel}

So far, we have built AGN obscuration by first randomly assigning a luminosity from the LC to any given AGN and then, based on their position within the LC, allocate a column density $N_{\rm H}$ according to the Models presented in Figure \ref{fig:models14}. We now explore a variant to this Evolution model in which we first compute the $N_{\rm H}$ column density {associated to each source based on their gas content and assumed gas geometry, allocate the source to a portion of the LC based on their retrieved value of $N_{\rm H}$, and then assign at random a bolometric luminosity to the SMBH within that portion of the LC.}
{For example, in Model 3, which is initially CTK, then CTN, and finally optical/UV visible, a galaxy with column density $\log N_{\rm H}/{\rm cm}^{-2} <22$ would be assigned a bolometric luminosity at random within the optical/UV visible portion of the LC, while a source with $\log N_{\rm H}/{\rm cm}^{-2}>24$ would have a luminosity selected from the initial CTK phase of the LC. Line-of-sight column densities $N_{\rm H}$ are calculated from the cold gas mass characterising each galaxy in GAEA, assuming an exponential gas density profile as detailed in \citetalias{Alonso-Tetilla} and summarised in Section \ref{subsec:hybridmodels} (we discuss in Section \ref{subsec:differentgasprofile} the impact of switching to a S\'{e}rsic geometry for the gas component). We call these Models ``hybrid'' because they combine Orientation and Evolution elements, in which the $N_{\rm H}$ is retrieved from a geometrical, line-of-sight framework, and the luminosity is assigned only within specific obscured or unobscured phases of the LC}. 

{In Figure \ref{fig:mixed34}, we show the hybrid variants applied to Evolution Models 3 and 4. We compare the fractions predicted by the Models with column densities extracted only from the host galaxy alone (dashed red lines), with those derived from the Models also inclusive of the fiducial torus component (solid blue lines)}. We find that the predicted fractions of CTK/CTN AGN are very similar in both models. We also note that the inclusion of a torus boosts, as expected, the AGN fractions in particular for CTK sources, which are otherwise difficult to generate in large quantities relying solely on the cold gas masses present in the host galaxies. All in all, the results in Figure \ref{fig:mixed34} align with the conclusions from \citetalias{Alonso-Tetilla}: the details of the shape of the LC or the exact luminosity assigned to an AGN within the LC, play a minor role in modulating the distributions of obscured AGN when $N_{\rm H}$ is calculated from a geometrical/orientation perspective as we already hinted in \citetalias{Alonso-Tetilla}, Appendix B. In \citetalias{Alonso-Tetilla} we in fact found that the degree of compactness and, to a lesser extent, the amount of cold gas in the host galaxy, are the main drivers in shaping the AGN fractions with luminosity {for AGN at cosmic noon}. 

\subsection{Redshift evolution} \label{sec:redshift}

In all our comparisons so far between model predictions and data we focused on a reference redshift of $z=2.4$, which is an epoch dominated by strong AGN and host galaxy star formation activity, and during which we would expect AGN moving from a more obscured phase to a more transparent one \citep[e.g.,][]{Granato+04, Alexander+05}. Here we aim to explore the predictions of our reference evolutionary models at other epochs, namely we choose $z=1$, and 3.3. We focus only on Model 3, noticing that similar results are found for the other models, not changing our conclusions. This model has been selected over the others since it presents the best match with \citetalias{Ueda+14, Ananna+19}. Figure \ref{fig:best-fits} reports the predictions of Model 3 without post-peak and without a torus but with a luminosity-dependent $\Delta\tau_{\rm QSO}$ (Eq. \ref{eq:lumdep2}, solid blue lines), and the reference Model 3 with a long-lived torus (dotted red lines). {For both Model 3 variants we find that their predicted fractions of CTN/CTK AGN are very similar within the redshift range explored here, in line with what also inferred in the data that suggest a very weak evolution. The long-lived torus Model 3 variant tends to predict gradually flatter obscured fractions of CTN AGN at later cosmic times. We ascribe this trend to the fact that when transitioning to lower redshifts, central SMBHs tend to experience longer post-peak phases and less accretion events, due to the less frequent mergers and less availability of cold gas, which both contribute to lower peak luminosities. In turn, this process decreases the probability of observing luminous AGN in the pre-peak (obscured) phase at late times, generating a reduction in the fraction of obscured luminous AGN, especially in the CTN regime. By contrast, the luminosity-dependent $\Delta\tau_{\rm QSO}$ Model 3 variant generates more stable and luminosity-dependent obscured fractions of AGN at all redshifts. However, the data are still too sparse to be able to clearly disfavour one Model variant from another from the redshift evolution alone}. 

\subsection{Can starbursts alone explain the fractions of obscured AGN at cosmic noon?} \label{sec:starburst}

\begin{figure}
    \centering
    \includegraphics[width=0.45\textwidth]{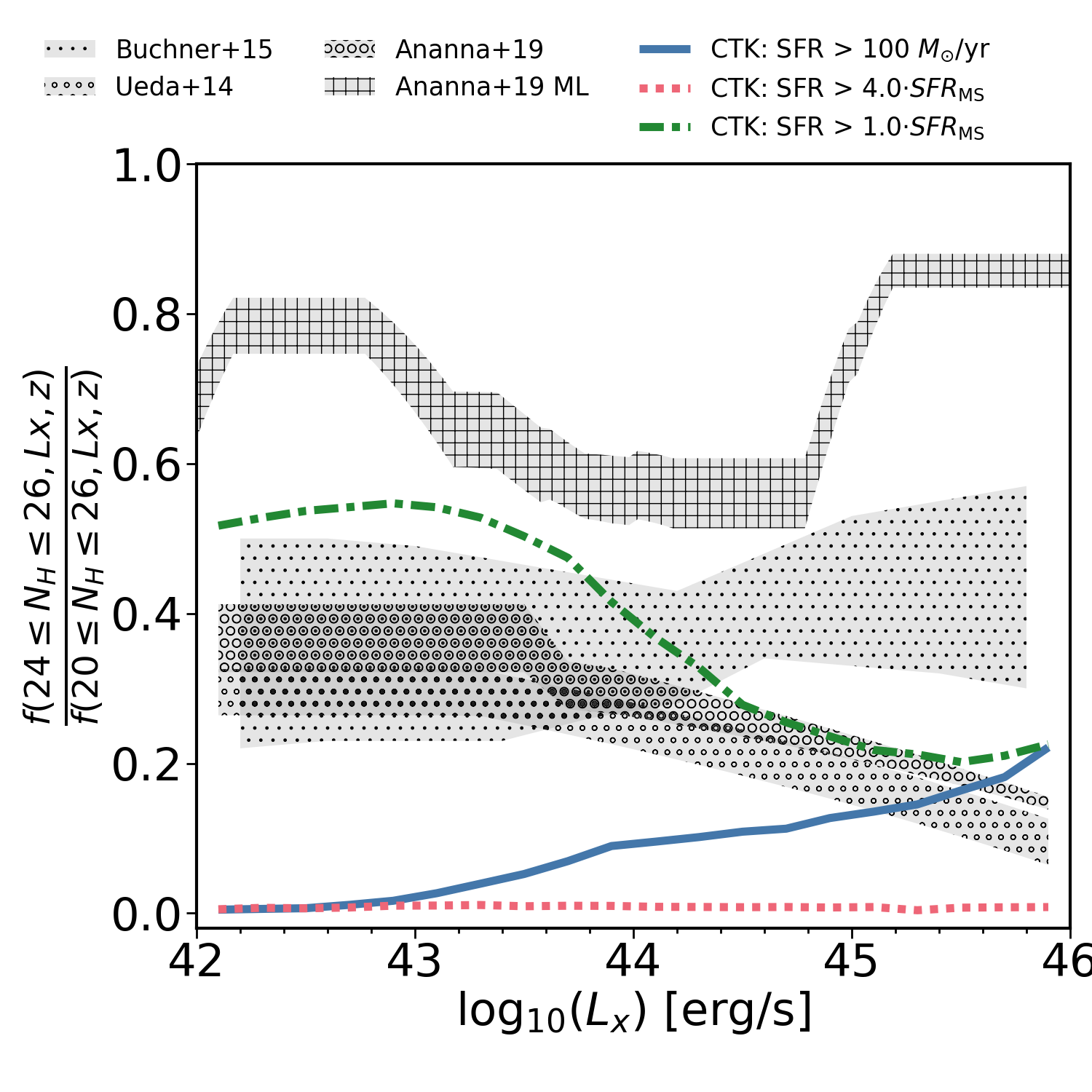}    
    \caption{Compton-thick (CTK) fractions of obscured AGN as a function of X-ray luminosity as predicted by models in which the CTK obscured galaxies are those selected above a certain threshold in star formation rate, as labelled. Only the model in which the CTK AGN are those with a star formation rate above the main sequence (MS; green, dot-dashed line) is able to generate a conspicous fraction of CTK AGN at all luminosities.}
    \label{fig:starburst}
\end{figure}

{In the previous Sections, we considered Evolution models built around the basic assumption that within the early growth of the central SMBH, the AGN/galaxy undergoes a CTN/CTK obscured phase. During the early phases of galaxy growth, especially for moderate/massive galaxies at $z>1$, the ones of interest to this work, one expects large gas reservoirs, along with intense star formation episodes coupled with proficient dust formation \citep[e.g,][]{Granato+06, Bate+22,Bosi25}. Therefore, many early starbursts could be associated with obscured AGN and, indeed, observations suggest a link between X-ray AGN activity and dust-enshrouded galaxies over a wide range of galaxy masses and redshifts \citep[e.g.,][]{Alexander+05,Mineo+12, Banerji+15, Bayliss+20, Carraro+20, Lim+20, Mountrichas+23, Riccio+23}.} 
{If starforming/starburst galaxies are located in the pre-peak phase of AGN activity \citep[e.g., Figure 7 in][]{Mountrichas+23}, we would expect that, irrespective of the details of the underlying AGN LC, all galaxies above a certain threshold of (specific) star formation rate, should be obscured up to a CTK level}. Indeed, as mentioned earlier, \citet{Andonie+23} recently found evidence for some highly starforming galaxies to be CTK AGN. Here we test whether GAEA galaxies characterised by a sufficiently large SFR could account for all CTK AGN at cosmic noon, without any reference to the underlying LC or gas fractions in the host, as expected in pure Evolution models \citep[e.g.,][]{Sanders+89, Hopkins+08}. 

\begin{figure*}
    \centering
    \includegraphics[width=0.9\textwidth]{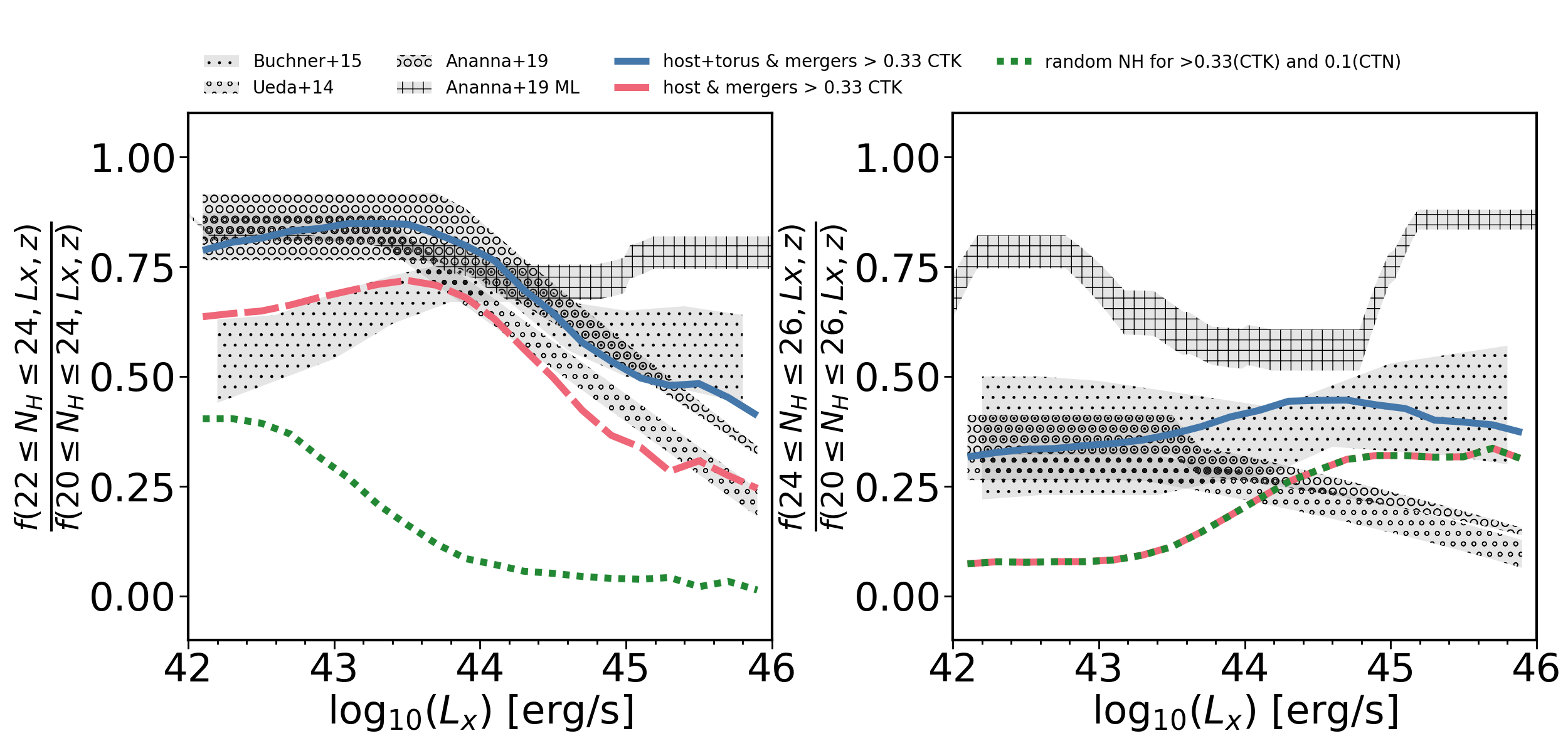}    
    \caption{{CTN and CTK AGN fractions as a function of X-ray luminosity as predicted by different merger obscuration models. The dashed red lines refere to our reference Orientation model, where all galaxies are characterised by a Hydrogen column density extracted from the gas exponential profile but where we also assume that all CTK AGN with $N_{\rm H}>10^{24}\, cm^{-2}$ originate from major mergers. The solid blue lines refere to a model variant identical to the previous one but also inclusive of our reference torus model. The dotted green lines finally show the predictions of a model where AGN triggered by mergers with a mass ratio between 0.1 and 1/3 are CTN and above 1/3 are CTK. The data are as in Figure \ref{fig:evolv12}. Models where CTK AGN are generated \textit{only} in major mergers seems to fall short, at least in GAEA, in reproducing the fractions of CTK AGN at lower luminosities}.}
    \label{fig:mergers}
\end{figure*}

{In Figure \ref{fig:starburst} we show the predicted fractions of CTK AGN on the assumption that these sources reside in host galaxies with SFRs \textit{i}) above 100 M$_{\odot}$/yr (solid blue line), \textit{ii}) four times above the (GAEA) main sequence (dotted red line), or \textit{iii}) just above the main sequence (dot-dashed green line). We note that, for simplicity, the AGN X-ray luminosities assigned to galaxies in Figure \ref{fig:starburst} are the peak luminosities in the LCs; assigning any other X-ray luminosity at random within the LC would yield very similar results, in line with what discussed in \citetalias{Alonso-Tetilla}.}

{The model assuming the $SFR>100\, M_{\odot}$/yr criterion (solid blue line), displays a steadily increasing CTK AGN fraction from $L_X\sim 10^{42-43}$ erg/s approaching the data only at the highest luminosities, in absolute constrast with any of the observational data. Very high SFR galaxies could still be CTK AGN \citep[e.g.,][]{Andonie+23}, but in our model those do not represent the bulk of the AGN population. The second model considering starbursts only those galaxies with $SFR>4\cdot SFR_{MS}$ (dotted red line), exhibits an almost flat and close to zero CTK fraction.  Our extreme model with all sources above the main sequence labelled as CTK AGN (dot-dashed green line), is the only one capable of generating a considerable fraction of CTK AGN at all luminosities, consistent, if not even slightly higher, than what inferred from the \citetalias{Ueda+14, Ananna+19} data. The results in Figure \ref{fig:starburst} therefore disfavour a model where starburst galaxies alone comprise the bulk of the CTK AGN at cosmic noon.} 

\subsection{The contribution of mergers to the fraction of obscured AGN} \label{sec:mergers}

As anticipated above, several models suggest that starbursts may be triggered by major mergers \citep[e.g.,][]{Mihos+96, DiMatteo+08, Zhou+18, Renaud+22}, and thus ultimately major mergers could be associated to obscured AGN \citep[e.g.,][]{Polletta+08, Riechers+13, Gilli+14, Ishibashi+16}. Indeed, in GAEA mergers trigger both accretion onto the central SMBH and an increase in the SFR. However, galaxies that experienced recent merger events do not necessarily position themselves above the main sequence \citep[e.g.,][]{Wang+19, Bluck+23, Blanquez-Sese+23}, although they could still be rich in gas and with an active central SMBH. Therefore, it could be argued that a major merger event, more than just a cut in SFR/sSFR, may be a more stringent tracer of a CTK AGN. In this Section we explore this possibility by selecting all galaxies in GAEA that had a recent major merger in the redshift of interest. More specifically, we select all {AGN} galaxies that have been triggered within 0.5 Gyr of the epoch of observation by a major merger with a progenitor mass ratio above 1/3 \citep[e.g.,][]{Stewart+09, Conselice+22}.

In Figure \ref{fig:mergers}, we show the results for three scenarios: 
\begin{itemize}
    \item {Scenario 1. We take our reference Orientation model, where all galaxies are characterised by a Hydrogen column density extracted from the gas exponential profile but where we also assume that all CTK AGN with $N_{\rm H} > 10^{24} {\rm cm}^{-2}$ originate from major mergers (dashed red line).}
    \item {Scenario 2. Our second scenario is identical to Scenario 1 but we also add the torus component (solid blue line).}
    \item {Scenario 3. Finally, building on the fact that $>90\%$ of our AGN at cosmic noon originate from mergers (see Section \ref{subsec:StarburstsMergerModels}), in our third scenario we assume that all AGN have a column density correlated to their merger status, more specifically those AGN triggered by a merger ratio below 1/10 have $N_{\rm H} \sim 10^{20}-10^{22} {\rm cm}^{-2}$, those with a ratio between $0.1$ and $1/3$ are CTN, and those with a ratio above 1/3 are CTK.}
\end{itemize}
As seen in the left panel of Figure \ref{fig:mergers}, the model in Scenario 1 (dashed red line), which assumes $N_{\rm H}$ values of the Orientation model for AGN not triggered by major mergers, yields a CTN fraction with a luminosity dependence similar to the observational data and to the one found in \citetalias{Alonso-Tetilla}. When the fiducial torus model is incorporated (Scenario 2, solid blue line), the fractions of CTN AGN increase as anticipated from the discussion in the previous Sections. However, the model in Scenario 3, which randomly assigns column densities to AGN only based on the merger ratio of their progenitors (dotted green line), is significantly below any of the data sets considered in this work. 
For the CTK fractions, plotted in the right panel of Figure \ref{fig:mergers}, scenarios 1 and 3 yield identical results, by design, as the CTK AGN only originate in major mergers, and generate CTK fractions that markedly drop to a few percent at faint luminosities below $\log L_X \lesssim 44$ erg/s (dashed red and dotted green lines). The model in scenario 2 instead, can generate significant fractions of CTK AGN even at faint luminosities due to the contribution of the torus.
All in all, a model where CTK AGN are generated only in major mergers seems to fall short, at least in GAEA, in reproducing the fractions of CTK AGN at lower luminosities, although major mergers alone may be sufficient to explain the large fractions at high luminosities, in line with some direct observations \citep[e.g.,][]{Treister+10, Treister+14, Gao+20, Tan+24, LaMarca25}.

\section{Discussion}\label{sec:discussion}

\subsection{The absolute number densities of AGN: inducing larger numbers of CTK AGN at the faint end with a torus component}
\label{subsec:AbsoluteNumbersAGN}


\begin{figure*}
    \centering
    \includegraphics[width=0.9\textwidth]{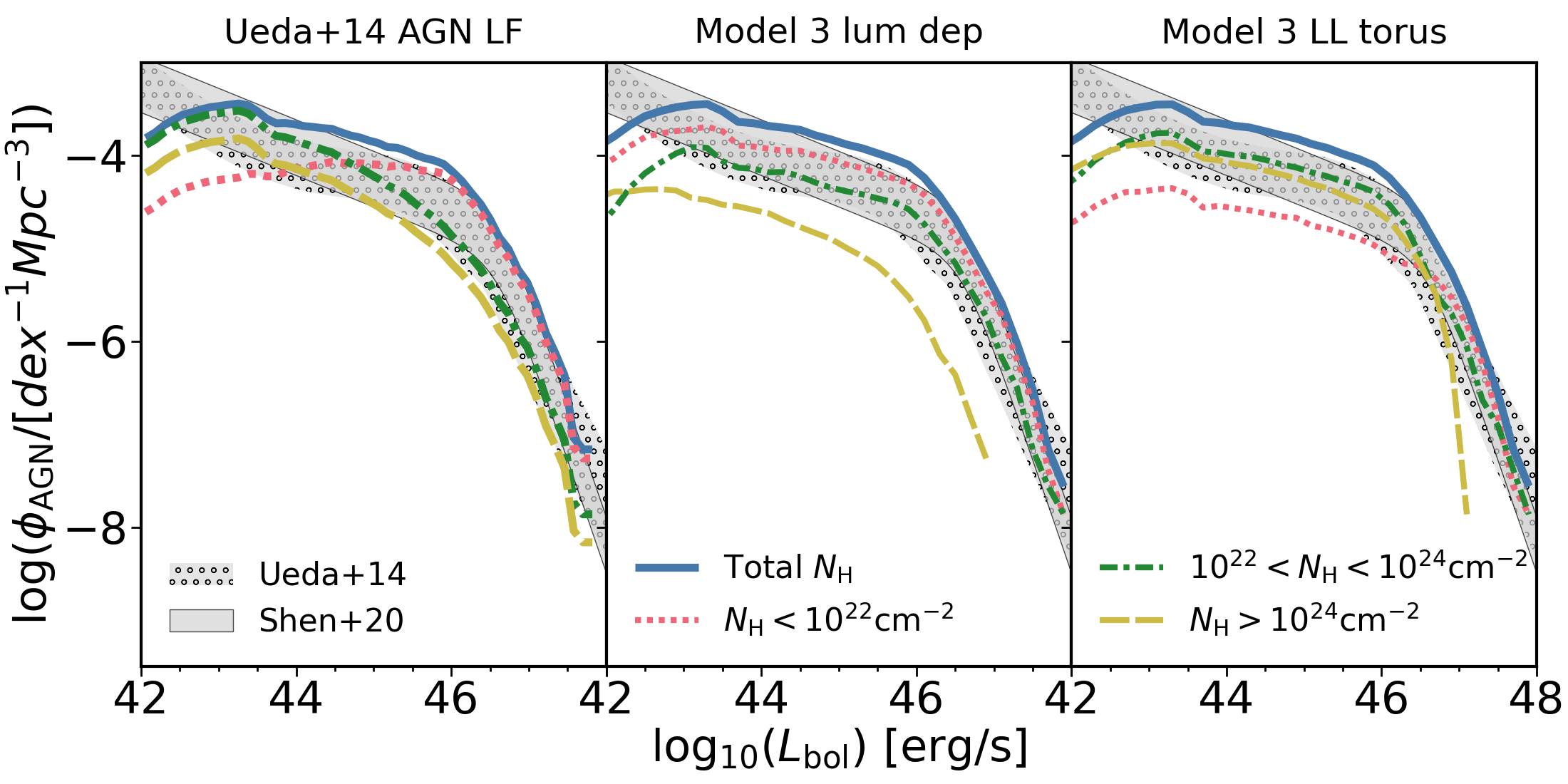}    
    \caption{{Left panel: Contributions to the total AGN bolometric luminosity function at $z=2.4$ from AGN of different $N_{\rm H}$ column densities, as labelled, using the model derived by \citetalias{Ueda+14}. Middle panel: Same format as the left panel but with predictions from Model 3 with a luminosity dependent optical/UV visibility window. Right panel: Same format as the middle panel with predictions from Model 3 with post-peak and the long-lived torus. The different lines refer to all AGN (solid blue line), CTN AGN (green dash-dotted line), CTK AGN (yellow dashed line) and optical/UV visible AGN (red dotted line). The observations on the total AGN bolometric luminosity function correspond to \citetalias{Ueda+14} (dotted area) and \citet{Shen+20} (shaded area), respectively. The variation of Model 3 with a torus generates a larger fraction of CTK AGN at all luminosities (right panel).}}
    \label{fig:AGNLF}
\end{figure*}

{So far, we have compared models to the fractions of obscured AGN as a function of X-ray luminosity, but we have not yet discussed the absolute number densities of AGN predicted by any model.
The central and right panels of Figure \ref{fig:AGNLF} show the total AGN luminosity function (LF) predicted by our reference Model 3 with, respectively, luminosity dependence in the input LC (Eqs. \ref{eq:prepeak} \& \ref{eq:postpeak}), and without luminosity dependence but with a long-lived torus. In each panel, the solid blue lines are the predicted total AGN LF (which, by design is the same for all models), while the dotted red, dot-dashed green, and long-dashed yellow lines trace, respectively, the optical/UV visible ($N_{\rm H} < 10^{22}$ cm$^{-2}$), CTN ($N_{\rm H} <10^{22}$ cm$^{-2}<N_{\rm H} < 10^{24}$ cm$^{-2}$), and CTK ($N_{\rm H} \ge 10^{24}$ cm$^{-2}$) contributions to the total AGN LF. As a comparison term, the left panel of Figure \ref{fig:AGNLF} reports the same total bolometric AGN LF predicted by our Model 3, but with contributions for AGN with different $N_{\rm H}$ cuts following the empirical model by \citetalias{Ueda+14}. As mentioned in Section \ref{sec:methodology} and further discussed by \citetalias{Fontanot+20}, our reference models broadly reproduce the observed AGN LF from \citet{Ueda+14, Shen+20}, which are included as grey regions in all panels of Figure \ref{fig:AGNLF}}. However, the distinct contributions from CTN and CTK sources to the total AGN LF vary noticeably when including a torus component in Model 3, as already {hinted at} in Figure \ref{fig:best-fits}. {In particular, the torus produces a boost in the fraction of CTK sources at all luminosities (right panel), and proportionally significantly reducing the number of unobscured AGN ($N_{\rm H} < 10^{22}$ cm$^{-2}$), whilst having a mild impact on the fraction of CTN AGN}. A torus component would thus tend to induce a predominance of CTN/CTK AGN below the knee of the AGN LF (dashed and dot-dashed lines in the right panel), a trend that can be directly tested in deep X-ray surveys. {Therefore, although the two renditions of Model 3, luminosity-dependent LC and added torus, may be roughly degenerate in the predicted CTN fractions, their absolute numbers of CTK AGN number densities could be very different and act as a powerful discriminators among successful models}.

\subsection{Dependence of the reference model on the peak luminosity of the light curve}
\label{subsec:DependenceOnLpeak}

{All the results on the fractions of obscured AGN presented in this work are based on the state-of-the-art semi-analytic model GAEA, which offers a self-consistent population of galaxies and their central SMBHs. In particular, in the GAEA model the peak luminosity $L_{\rm peak}$, located at $t_{\rm peak}$, the time at which the central SMBH ends its exponential growth phase, is reached when the accretion rate reaches its maximum value (Eq. \ref{eq:mcrit}), as suggested by tailored hydrodynamic simulations (see Section \ref{subsec:LCs}). Notably, we have verified that randomly shifting $t_{\rm peak}$ within the LC, that is, without imposing any link between the peak and critical mass accretion rates, does not affect the predicted relative fractions of obscured AGN. The reason for such a weak dependence is that $t_{\rm peak}$ distribution is already quite uniformly distributed within the AGN LC, thus a random reshuffling has a negligible impact on the model predictions. Increasing or decreasing $L_{\rm peak}$ instead moves proportionally the AGN LF towards brighter or fainter luminosities, respectively, without significantly impacting its shape nor the implied AGN obscured fractions. These tests further reinforce the conclusion that, within the context of Evolution models, the main drivers behind the fractions of obscured AGN are not to be searched within the specific shape of the underlying LC, but rather in the relative portions of the LC that are obscured versus those that are unobscured.}

\subsection{Comparing inferred optical/UV visibility windows with independent estimates from AGN clustering}\label{subsec:clustering}

{In Section \ref{sec:results} we showed that to induce a noticeable luminosity dependence in the fractions of obscured AGN, decreasing with increasing AGN luminosity, we could either allow intrinsically more luminous AGN to be visible in the optical/UV for a longer period than fainter AGN, i.e., increasing their visibility window $\Delta \tau_{\rm QSO}$, or invoke a long-lived torus component, the thickness of which decreases with AGN power}. A long-lived torus does not rule out the possibility of a torus life cycle, which is destroyed and reformed during several AGN periods \citep[as suggested by, e.g.,][]{Garcia-Burillo+19, Garcia-Burillo+21, GarciaBurillo24}. {Irrespective of the details of the specific assumption adopted, all our models point to an optical/UV visibility window in the range $\Delta \tau_{\rm QSO} \sim 10^7 - 10^8$ yr, reaching $\Delta \tau_{\rm QSO} \sim 10^8$ yr for the most luminous AGN (see further details in Section \ref{subsec:purevolution}).} 

{On the assumption of a background dark matter-dominated Universe, clustering analysis can provide  relevant insight on the distribution of host dark matter haloes, in terms of spatial distribution and number densities, which, in turn, when compared with the observed number densities of AGN, can yield valuable and independent constraints on the AGN duty cycles and LCs \citep[e.g.,][]{Richstone+98,HaimanHui,MartiniWeinberg,Martini+04,Hopkins07clustering,Croton09,Hennawi10,Shankar10clustering,Shen10clustering}. The quasar clustering analysis from \citet{Eilers+24} points to short QSO visibility windows in the range $10^5-10^7$ yr.  Via a semi-empirical model of QSOs built around the observational results from \citet{Eilers+24}, \citet{Pizzati+24} suggested a strong evolution of the duty cycle, approaching $10^8$ yr at $z\sim2-3$, a value consistent with what inferred from our reference models, especially for the most luminous sources. 
}

{For completeness, we have also computed that the 2-point correlation function predicted by our hybrid model at $z=2.4$ and found consistency with the large-scale $r_p>1 $ Mpc clustering strength measured by \citet{Viitanen+23} for X-ray selected AGN in the range $1.1<z<3$, at least for galaxies with stellar mass $M_{\star} > 10^{10.5} \, M_{\odot}$. We also found that the AGN clustering signal predicted by our reference models at large scales is independent of the column density of the host galaxy or the inclusion or not of a torus-like component in the model.}

\subsection{The effect of alternative gas density profiles in Orientation models and their relation with galactic compactness}
\label{subsec:differentgasprofile}

{In our hybrid models, the column densities are computed on a galaxy-by-galaxy basis from the gas content in the host galaxy, assumed to follow an exponential density profile with a scale radius smaller than the one characterising the stellar component. More precisely, in \citetalias{Alonso-Tetilla}, we introduced the compactness parameter $N$ as the ratio between the gas scale length and the stellar scale length, $R_{\rm d, gas} = N \cdot R_{\rm d, \star}$, and found that $N \sim 0.3$ is a sufficient condition to achieve a CTN level of obscuration comparable to observational data. Interestingly, the preferred value of $N$ adopted in our reference models is in line with independent observational results by, e.g., \citet{Puglisi+21}, \citet{Liu+24}, \citet{Price25}, who found evidence from combined ALMA and Herschel data that, on average, $R_{\rm d, gas} \sim 0.3 \cdot R_{\rm d, \star}$ in main-sequence galaxies at cosmic noon. These findings suggest that obscuration arising solely from the large-scale interstellar medium within the host galaxy may be insufficient, at least at $z<3$, to account for the large fractions of highly obscured AGN, pointing to the need for an additional obscuring component, such as a torus, to produce most of CTK obscuration in galaxies.}

As discussed in \citetalias{Alonso-Tetilla}, this conclusion was based on the strict assumption of forcing an exponential profile on the gas component in all host galaxies. However, there is also some evidence for more compact/complex morphologies defining starforming and dust-enshrouded galaxies at higher redshifts \citep[e.g.,][]{Tan+24, Hodge+24}. {To this purpose, we have also explored the implications of switching to a S\'{e}rsic gas density profile for our galaxies under the Orientation model}. 

We find that when implementing a S\'{e}rsic profile whilst maintaining a compactness ratio $N=R_{\rm d,gas}/R_{\rm d,\star}\sim0.3$, similar to what suggested in observations \citep[][]{Puglisi+19, Puglisi+21}, the models tend to generate very large saturated fractions of CTN AGN close to unity, irrespective of the S\'{e}rsic index chosen. A S\'{e}rsic profile also tends to generate a larger fraction of CTK sources, as discussed above, introducing a potential degeneracy between the torus and the geometry of the galaxy, complicating the distinction between torus-driven and host-galaxy-driven obscuration. {Targetted deep observations via, e.g., JWST, will be crucial in providing direct evidence of the large-scale obscuring properties of galaxies at high redshift. For example, \citet{Silverman+23} using JWST/NIRCam imaging from the COSMOS-Web survey, probed the galaxy-wide dust distribution of a few X-ray AGN up to $z \sim 2$ finding evidence for an average $N_{\rm H} \sim 10^{22.5}$ cm$^{-2}$, well within the CTN regime}. 


{To re-establish a good alignment with the observed CTN fractions, in particular with the measurements by \citetalias{Ueda+14}, maintaining a S\'{e}rsic gas density profile throughout, we would require the compactness ratio to decrease to $N\sim0.01-0.1$, which are much lower values than what is estimated in hydrodynamic simulations and dedicated observations \citep[e.g.,][and references therein]{Puglisi+19, Puglisi+21,Price25}}. We thus retain the exponential profile for the gas component in our AGN hosts at $z<3$. Indeed, \citet[][see also \citealt{SmetJWSTsizes}]{Lyu+24} find that starforming galaxies at $z<2.5$, well characterized by an exponential profile for the stellar component, also have a more compact starforming disc close to exponential \citep[e.g.,][]{Magnelli+23, Shen+23}, consistent with what expected from the wet compaction scenario \citep{Tacchella+15, Barro+17, Lapiner+23}. \citet{Lyu+24} also find that a high percentage of massive galaxies ($\sim 30\%$ of $M_{\star} > 10^{10.5} M_{\odot}$) have compact star-forming cores, while \citet{Puglisi+21} report even a higher percentage of $\sim 50\%$ for $M_{\star} > 10^{11} M_{\odot}$ \citep[see also][]{Pozzi+24}.

\subsection{The implications of missing obscured AGN populations such as Little Red Dots}
\label{subsec:LRDs}

Recent studies of Little Red Dots (LRDs), facilitated by JWST observations \citep[e.g.,][]{Akins+23, Akins+24, Matthee+24, Perez-Gonzalez+24, Polletta+24, Durodola+24}, have been identified with extremely compact and highly obscured galaxies but invisible in X-ray wavelengths, and therefore not in our catalogues. LRDs could potentially provide an additional source of CTK AGN, although it is not clear if these sources are intrinsically X-ray weak \citep[e.g.,][]{Maiolino+24}, or even fully-fledged obscured AGN \citep[e.g.,][]{Baggen+24}. {They are mostly located at $z>4$ according to current observations \citep[e.g.,][]{Akins+23, Iani+24, Kokorev+24, BisigelloLRDs, Ma25LRDs}, and would not be recorded in the AGN LFs or obscured fractions based on the X-ray data that we are using as a reference to calibrate the models in this work. We have thus not considered them explicitly in this work. In addition, there are other populations of obscured sources like changing-look AGN \citep[e.g.,][]{RicciTrakhtenbrot23}, optically quiescent quasars \citet[][]{Greenwell+24}, galaxies with a large degree of clumpiness in their gas disc \citep[e.g.,][]{Gilli+22}, that we might be missing in our comparison data.} We note, however, that even allowing for a significantly larger fraction of CTK AGN in our reference data, also hinted by the \citetalias{Ananna+19}-ML results, would just strengthen our conclusions on the need for a long-lived torus-like component, or in pure Evolution models, by a more extended CTK phase, generated by either a brief post-peak phase, as in Model 3, or a more extended CTK post-peak phase, as in Model 4. Indeed, \citet{Maiolino+24} discussed the concrete possibility for many high-$z$ LRDs to be CTK sources with obscuration arising from central dust-free CTK clouds. 

\subsection{Comparison with previous works}

Previous models provide valuable context for our findings. Our Evolution models bear similarities to those by \citet{DiMatteo+05}, which predict a specific duration for the luminous episode of a self-regulated SMBH. It typically lasts until the feedback energy expels enough gas to significantly reduce the accretion rate, effectively ending the quasar phase, producing a very short or negligible post-peak phase, similar to our Model 3. While the exact duration can vary depending on the specific conditions of each galaxy merger, the simulations provide a framework to predict that this luminous episode is relatively short and self-limiting due to the feedback processes. Despite our LC featuring a longer pre-peak evolution and a more rapid post-peak decline than their model, we find similar LC results. Additionally, \citet{Georgantopoulos+23} suggest that unobscured AGN tend to reside in younger galaxies, whereas obscured AGN are found in galaxies between the young and old population stages. These interesting results would challenge both traditional Orientation and Evolution models, but would possibly be more aligned with our Model 4, where obscuration could still occur even during later phases of the peak AGN activity. On the other hand, \citet{Parlanti+23} used JWST images to demonstrate that dust-obscured galaxies represent an evolutionary stage preceding the unobscured quasar phase. This finding aligns well with the traditional Evolution model, but could still be accommodated within our Model 4 which allows for both a brief visible QSO phase and then a more obscured one at a larger stage. {From pure continuity evolution models that self-consistently grow the population of SMBHs via the input AGN bolometric luminosity function, \citet{Aversa+15} showed that their preferred LCs are those without an extended post-peak phase, in line with our Model 3 without a post-peak phase which better reproduce the high fractions of obscured AGN. Similar results on sharp drops post-peak in the AGN LCs have also been put forward by \citet{Lapi+14} when modelling the SMBH-galaxy coevolution combining FIR, X-ray, and optical/UV data.} 

\section{Conclusions} \label{sec:conclusions}

{The sources of obscuration in AGN remain debated, as they may originate from large-scale obscuration within the galaxy, which could vary in time and space, as well as from an inner dusty torus component surrounding the central SMBH}. In this paper, we have modelled the obscuration in AGN using a comprehensive, {combined semi-analytic and semi-empirical framework that incorporates various elements from Evolution models, most notably an evolving Hydrogen column density $N_H$ varying within the AGN light curve (LC), also combined with key Orientation components, such as an inner torus and obscuration arising from the cold gas in the interstellar medium}. Additionally, we have investigated the impact of starburst activity and mergers on obscuration. Our main results can be summarized as follows:

\begin{itemize}
\item Traditional Evolution models characterized by LCs with a long post-peak phase and CTN/CTK obscuration only pre-peak, tend to struggle in reproducing the observed large fractions of CTN/CTK X-ray AGN at $1<z<3$. {These findings align with those feedback models suggesting that AGN outflows may not be as efficient in removing gas in galaxies \citep[][]{Menci+2019}}.
\item Evolution models characterized by LCs with a sharp drop after the peak (Models 2 and 3), or with a CTN phase reappearing after the peak (Model 4), are favoured in reproducing the high fractions of CTN/CTK AGN.
\item A steep drop in the obscured fractions of AGN as observed in some data sets can be reproduced in our pure Evolution models by including a luminosity dependence in the UV/optical visibility window $\Delta \tau_{\rm QSO}$, increasing from $\Delta \tau_{\rm QSO} = 10^7$ yr to $\Delta \tau_{\rm QSO} = 8\cdot10^7$ yr for the more luminous sources.
\item All our models tend to align with the fractions of obscured X-ray AGN when the visibility window is of the order of \(\Delta \tau_{\rm QSO} \sim 10^7 - 10^8\) yr. This range is broadly consistent with the latest estimates extracted from QSO 2-point correlation functions. 
\item Including in Evolution models a torus component living throughout the AGN light curve, surviving beyond the peak of AGN's energetic output, and with a thickness decreasing with increasing AGN power, can increase the fractions of CTN and CTK AGN whilst also inducing a significant luminosity dependence {in line with some data dets}. An inner geometrical component like a torus would imply that orientation effects tend to constitute a key source of obscuration even in Evolution models {at cosmic noon}. 
\item We find that the fraction of sources in the model observationally defined as starbursts lying four times above the MS, falls short in reproducing the fractions of CTK AGN. {Similarly, also the number of major mergers with mass ratio $>1/3$ tends to be too low to account for the significant fractions of CTK AGN, except possibly for the most luminous AGN, suggesting that either a torus-like component is ubiquitous in all types of AGN, and/or that large-scale obscuration is triggered in galaxies by in-situ large column densities and/or compact morphologies}. 
\end{itemize}

{In summary, our current results suggest that the traditional view of Evolution models may still be valid but under non-trivial fine-tuning conditions and possibly supported by a central, long-lived torus-like component playing a pivotal role in shaping obscuration in AGN. Our predictions can be tested with current and new facilities such as JWST, ALMA or NewAthena.}

\section*{Acknowledgements}

{We warmly thank the referee for a thorough and insightful review that helped us to significantly improve the overall presentation of the paper and highlight our main results.} This study has been carried within BiD4BESt \footnote{More information about BiD4BESt and the Innovative Training Network can be found in \url{https://www.bid4best.org/}.}, an Innovative Training Network (ITN) providing doctoral training in the formation of supermassive black holes in a cosmological context. BiD4BESt has received funding from the European Union’s Horizon 2020 research and innovation programme under the Marie Skłodowska-Curie grant agreement No 860744 (grant coordinator F. Shankar). 
AL is partly supported by the PRIN MIUR 2017 prot. 20173ML3WW 002 ``Opening the ALMA window on the cosmic evolution of gas, stars, and massive black holes''. 
FF and NM acknowledge support from PRIN MIUR project ``Black Hole winds and the Baryon Life Cycle of Galaxies: the stone-guest at the galaxy evolution supper'', contract 2017-PH3WAT. 
AP acknowledges partial support by STFC through grants ST/T000244/1 and ST/P000541/1.
CRA acknowledges support from the Agencia Estatal de Investigaci\'on of the Ministerio de Ciencia, Innovaci\'on y Universidades (MCIU/AEI) under the grant ``Tracking active galactic nuclei feedback from parsec to kiloparsec scales'', with reference PID2022-141105NB-I00 and the European Regional Development Fund (ERDF).
MV is supported by the Italian Research Center on High Performance Computing, Big Data and Quantum Computing (ICSC), project funded by European Union - NextGenerationEU - and National Recovery and Resilience Plan (NRRP) - Mission 4 Component 2, within the activities of Spoke 3, Astrophysics and Cosmos Observations, and by the INFN Indark Grant.

This research made use of \textit{Numpy} \citep[][]{Harris+2020}, and \textit{Scipy} \citep[][]{Virtanen+2020} modules. We thank the arXiv for providing a preprint server that has helped to facilitate the dissemination of our results. This research has made use of the NASA/IPAC Extragalactic Database (NED), which is operated by the Jet Propulsion Laboratory, California Institute of Technology, under contract with the National Aeronautics and Space Administration.

\section*{Data Availability}

An introduction to {\sc gaea}, a list of our recent work, as well as datafiles containing published model predictions, can be found at \url{https://sites.google.com/inaf.it/gaea/home}. The analysis carried out in this work as well as the plotted results will be available upon request until a free access database is released (which will be found in \url{https://github.com/AVAlonso}).



\bibliographystyle{mnras}
\bibliography{example} 





\bsp	
\label{lastpage}
\end{document}